\documentclass{IEEEtaes}

\usepackage{color,array,amsthm}
\usepackage{amssymb}
\usepackage{graphicx}
\usepackage{soul}
\usepackage{pdflscape} 
\usepackage{graphicx} 
\usepackage{makecell} 
\usepackage{float}
\usepackage{booktabs}
\usepackage{lipsum}
\usepackage{multicol}%
\usepackage{rotating}
\usepackage{ amssymb}
\usepackage{afterpage}
\usepackage{subfigure}
\usepackage{academicons}
\usepackage{orcidlink}
\usepackage{multirow}
\jvol{XX}
\jnum{XX}
\jmonth{XXXXX}
\paper{1234567}
\pubyear{2026}
\doiinfo{TAES.2020.Doi Number}

\usepackage[acronym]{glossaries}
\usepackage{graphicx}

\newacronym{AD}{AD}{Angular Distance}
\newacronym{ADC}{ADC}{Analog to Digital Converter}
\newacronym{AWGN}{AWGN}{Additive White Gaussian Noise}
\newacronym{AGN}{AGN}{Additive Gaussian Noise}
\newacronym{AMSS}{AMSS}{Angular Metric for Shape Similarity}

\newacronym{AGNSS}{AGNSS}{Aided GNSS}
\newacronym{AI}{AI}{Artificial Intelligence}
\newacronym{AS}{AS}{Angular Similarity}
\newacronym{ASI}{ASI}{Agenzia Spaziale Italiana}
\newacronym{AT}{AT}{Aiding Trajectory}
\newacronym{ATD}{ATD}{Accumulated Time Difference}
\newacronym{BD}{BD}{Binary Distance}
\newacronym{BDS}{BDS}{BeiDou Navigation Satellite System}
\newacronym{BGM1}{BGM1}{Blue Ghost Mission 1}
\newacronym{BLUE}{BLUE}{Best Unbiased Linear Estimator}
\newacronym{BOC}{BOC}{Binary Offset Code}
\newacronym{BPSK}{BPSK}{Binary Phase-Shifting Keying}
\newacronym{C/A}{C/A}{Coarse/Acquisition}
\newacronym{CDSN}{CDSN}{Chinese Deep Space Network}
\newacronym{CLPS}{CLPS}{Commercial Lunar Payload Services}
\newacronym{CanX-2}{CanX-2}{Canadian Advanced Nanosatellite eXperiment-2}
\newacronym{CAF}{CAF}{Cross Ambiguity Function}
\newacronym{CEP}{CEP}{Circular Error Probable}
\newacronym{CDF}{CDF}{Cumulative Distribution Function}
\newacronym{CDMA}{CDMA}{Code Division Multiple Access}
\newacronym{C/N0}{$C/N_0$}{Carrier to Noise Density Ratio}
\newacronym{CRC}{CRC}{Cyclic Redundancy Check}
\newacronym{CW}{CW}{Continuous Wave}
\newacronym{NB}{NB}{Narrow Band}
\newacronym{CS}{CS}{Cosine Similarity}
\newacronym{DC}{DC}{Direct Current}
\newacronym{DF}{DF}{Distance Function}
\newacronym{DFT}{DFT}{Discrete Fourier Transform}
\newacronym{DLL}{DLL}{Delay Lock Loop}
\newacronym{DoD}{DoD}{Department of Defense}
\newacronym{DSN}{DSN}{Deep Space Network}
\newacronym{DTE}{DTE}{Direct-to-Earth}
\newacronym{DTW}{DTW}{Dynamic Time Warping}
\newacronym{DDTW}{DDTW}{Derivative Dynamic Time Warping}
\newacronym{DGNSS}{DGNSS}{Differential GNSS}
\newacronym{DOP}{DOP}{Dilution Of Precision}
\newacronym{ECDF}{ECDF}{Empirical Cumulative Distribution Function}
\newacronym{ECEF}{ECEF}{Earth-Centered Earth-Fixed}
\newacronym{EDR}{EDR}{Edit Distance on Real sequences}
\newacronym{EGNOS}{EGNOS}{European Geostationary Navigation Overlay Service}
\newacronym{EKF}{EKF}{Extended Kalman Filter}
\newacronym{ENU}{ENU}{East-North-Up}
\newacronym{ERP}{ERP}{Edit distance on Real Penalty}
\newacronym{ESA}{ESA}{European Space Agency}
\newacronym{ESTRACK}{ESTRACK}{European Space Tracking Network}
\newacronym{EU}{EU}{European Union}
\newacronym{ED}{ED}{Euclidean Distance}
\newacronym{EUSPA}{EUSPA}{European Union Agency for the Space Programme}
\newacronym{FDMA}{FDMA}{Frequency Division Multiple Access}
\newacronym{GAGAN}{GAGAN}{GPS-Aided GEO Augmented Navigation}
\newacronym{GEONS}{GEONS}{Goddard Enhanced Onboard Navigation System}
\newacronym{GGMS}{GGMS}{GEONS Ground Matlab Simulation}
\newacronym{FFT}{FFT}{Fast Fourier Transform}
\newacronym{FLL}{FLL}{Frequency Lock Loop}
\newacronym{GDOP}{GDOP}{Geometric Dilution Of Precision}
\newacronym{GEO}{GEO}{Geostationary Earth Orbit}
\newacronym{GLONASS}{GLONASS}{GLObal'naya NAvigatsionnaya Sputnikovaya Sistema}
\newacronym{GNC}{GNC}{Guidance Navigation and Control}
\newacronym{GNSS}{GNSS}{Global Navigation Satellite System}
\newacronym{GPS}{GPS}{Global Positioning System}
\newacronym{GS}{GS}{Ground Stations}
\newacronym{HAS}{HAS}{High Accuracy Service}
\newacronym{HEO}{HEO}{High Earth Orbit}
\newacronym{HDOP}{HDOP}{Horizontal Dilution Of Precision}
\newacronym{HGA}{HGA}{High Gain Antenna}
\newacronym{HIL}{HIL}{Hrdware In the Loop}
\newacronym{ICD}{ICD}{Interface Control Documents}
\newacronym{IRNSS}{IRNSS}{Indian Regional Navigation Satellite System}
\newacronym{IDFT}{IDFT}{Inverse Discrete Fourier Transform}
\newacronym{ISTRAC}{ISTRAC}{ISRO Telemetry, Tracking and Command Network}
\newacronym{IGS}{IGS}{International GNSS Service}
\newacronym{IQS}{IQS}{In-phase and Quadrature Samples}
\newacronym{IQ}{IQ}{In-Phase Quadrature}
\newacronym{IF}{IF}{Intermediate Frequency}
\newacronym{IGSO}{IGSO}{Inclined Geosynchronous Orbit}
\newacronym{LCSS}{LCSS}{Longest Common Sub Sequence}
\newacronym{JAXA}{JAXA}{Japan Aerospace Exploration Agency}
\newacronym{JPL}{JPL}{Jet Propulsion Laboratory}
\newacronym{KF}{KF}{Kalman Filter}
\newacronym{LuGRE}{LuGRE}{Lunar GNSS Receiver Experiment}
\newacronym{LEO}{LEO}{Low Earth Orbit}
\newacronym{LLA}{LLA}{Latitude, Longitude, Altitude}
\newacronym{LLO}{LLO}{Low Lunar Orbit}
\newacronym{LOS}{LOS}{Line of Sight}
\newacronym{LMS}{LMS}{Least Mean Square}
\newacronym{LNA}{LNA}{Low Noise Amplifier}
\newacronym{ML}{ML}{Maximum Likelihood}
\newacronym{MMS}{MMS}{Magnetospheric Multiscale}
\newacronym{MC}{MC}{Monte Carlo}
\newacronym{MD}{MD}{Manhattan Distance}
\newacronym{MDMS}{MDMS}{Mono Dimensional Matching Search}
\newacronym{MDS}{MDS}{Multi-Dimensional Sequence}
\newacronym{MEO}{MEO}{Medium Earth Orbit}
\newacronym{MM}{MM}{Moment Method}
\newacronym{MS}{MS}{Moon Surface}
\newacronym{MTO}{MTO}{Moon Transfer Orbit}
\newacronym{NASA}{NASA}{National Aeronautics and Space Administration}
\newacronym{NPDI}{NPDI}{Non-coherent Post Detection Integration}
\newacronym{NavIC}{NavIC}{Navigation Indian Constellation}
\newacronym{NavSAS}{NavSAS}{Navigation Signal Analysis and Simulation}
\newacronym{NLOS}{NLOS}{Non-Line-Of-Sight}
\newacronym{NWPR}{NWPR}{Narrow-Wideband Power Ratio}
\newacronym{OD}{OD}{Orbit Determination}
\newacronym{OF}{OF}{Orbital Filter}
\newacronym{OS}{OS}{Open Service}
\newacronym{PDF}{PDF}{Probability Density Function}
\newacronym{PLL}{PLL}{Phase Lock Loop}
\newacronym{PMF}{PMF}{Probability Mass Function}
\newacronym{POD}{POD}{Precise orbit determination}
\newacronym{PRN}{PRN}{Pseudo Random Noise}
\newacronym{PNT}{PNT}{Position, Navigation, Timing}
\newacronym{PPP}{PPP}{Precise Point Position}
\newacronym{PRS}{PRS}{Public Regulated Service}
\newacronym{PSD}{PSD}{Power Spectral Density}
\newacronym{PVT}{PVT}{Position, Velocity, Time}
\newacronym{QMBOC}{QMBOC}{Quadrature Multiplexed Binary Offset Carrier}
\newacronym{QZO}{QZO}{Quasi-Zenith Orbit}
\newacronym{QZSS}{QZSS}{Quasi-Zenith Satellite System}
\newacronym{R}{R}{Correlation}
\newacronym{RAAN}{RAAN}{Right Ascension of the Ascending Node}
\newacronym{RNSS}{RNSS}{Regional Navigation Satellite System}
\newacronym{RMSE}{RMSE}{Root Mean Squared Error}
\newacronym{RTK}{RTK}{Real Time Kinematic}
\newacronym{RE}{RE}{Earth Radii}
\newacronym{RF}{RF}{Radio Frequency}
\newacronym{RFCS}{RFCS}{Radio Frequency Constellation Simulator}
\newacronym{ROC}{ROC}{Receiver Operating Characteristic}
\newacronym{SAR}{SAR}{Search and Rescue Service}
\newacronym{SBAS}{SBAS}{Satellite-based Augmentation System}
\newacronym{SC}{SC}{Sample Capture}
\newacronym{SDR}{SDR}{Software Defined Radio}
\newacronym{SIS}{SIS}{Signal in Space}
\newacronym{SM}{SM}{Sequence matching}
\newacronym{SNV}{SNV}{Squared Signal-to-Noise Variance}
\newacronym{SV}{SV}{Satellite Vehicle}
\newacronym{SSDTW}{SSDTW}{Sub-Sequence Dynamic Time Warping}
\newacronym{SSV}{SSV}{Space Service Volume}
\newacronym{TA-EKF}{TA-EKF}{Trajectory Aware - Extended Kalman Filter}
\newacronym{TEC}{TEC}{Total Electron Content}
\newacronym{ToA}{ToA}{Time of Arrival}
\newacronym{ToF}{ToF}{Time of Flight}
\newacronym{TXCO}{TXCO}{Temperature Compensated Crystal Oscillator}
\newacronym{UERE}{UERE}{User Equivalent Range Error}
\newacronym{US}{US}{United States}
\newacronym{UTC}{UTC}{Coordinated Universal Time}
\newacronym{VDMDS}{VDMDS}{Variable-Dimension Multi Dimensional Sequence}
\newacronym{WGDOP}{WGDOP}{Weighted Geometric Dilution Of Precision}
\newacronym{WLMS}{WLMS}{Weighted Least Mean Square}
\newacronym{WLS}{WLS}{Weighted Least Square}
\newacronym{WP}{WP}{Warping Path}
\newacronym{XO}{XO}{Crystal Oscillator}
\newacronym{RTP}{RTP}{Real-Time Processing}

\begin{document}

\title{First Multi-Constellation Observations of Navigation Satellite Signals in the Lunar Domain by Post-Processing L1/L5 IQ Snapshots} 








\author{LORENZO SCIACCA\,\orcidlink{0009-0007-6752-3032}}
\member{Graduate Student Member, IEEE}

\author{ALEX MINETTO\,\orcidlink{0000-0002-0586-7151}}
\member{Member, IEEE} 

\author{ANDREA NARDIN\,\orcidlink{0000-0001-9167-2272}}
\member{Member, IEEE}

\author{FABIO DOVIS\,\orcidlink{0000-0001-6078-9099}}
\member{Member, IEEE}
\affil{Politecnico di Torino, Turin, Italy} 
\author{LUCA CANZIAN}
\affil{Qascom Srl, Cassola, Italy}

\author{MARIO MUSMECI\,}
\author{CLAUDIA FACCHINETTI\,}
\author{GIANCARLO VARACALLI\,}
\affil{Agenzia Spaziale Italiana, Rome, Italy}


\receiveddate{Manuscript received XXXXX 00, 0000; revised XXXXX 00, 0000; accepted XXXXX 00, 0000.\\
DOI. No. XXXXX\\
Refereeing of this contribution was handled by XXXX XXXX.
This study was funded within the contract n. 2021-26-HH.0 between Agenzia Spaziale Italiana and Politecnico di Torino
”Attività di Ricerca e Sviluppo inerente alla Navigazione GNSS nello Space volume Terra/Luna nell’ambito del Lunar GNSS
Receiver Experiment”.
This publication is also part of the project PNRR-NGEU which has received funding from the MUR – DM 630/2024.
 }


\authoraddress{Authors’ addresses: Lorenzo Sciacca, Alex Minetto, Andrea Nardin, Fabio Dovis are with the Department of Electronics and
Telecommunications, Politecnico di Torino, Turin, TO, 10124 Italy (e-mail:
name.surname@polito.it); Luca Canzian is with Qascom Srl, Cassola, VI, 36022 Italy (e-mail: luca.canzian@qascom.it)}


\markboth{SCIACCA ET AL.}{First Multi-Constellation Observations of Navigation Satellite signals in the Lunar Domain}
\maketitle

\begin{abstract}

The use of Global Navigation Satellite Systems (GNSS) to increase spacecraft autonomy for orbit determination has gained renewed momentum following the Lunar GNSS Receiver Experiment (LuGRE), which demonstrated feasible onboard GPS and Galileo signal reception and tracking at lunar distances. This work processes in-phase and quadrature (IQ) snapshots collected by the LuGRE receiver in cis-lunar space and on the lunar surface to assess multi-frequency, multi-constellation signal availability. 
Signals from additional systems beyond GPS and Galileo, including RNSS and SBAS constellations, are observable and successfully acquired exclusively in the recorded IQ snapshots.
These observations provide the first experimental evidence that signals from multiple constellations—including systems not supported by LuGRE real-time operations—are detectable at unprecedented distances from Earth. Useful observables can be extracted from the IQ snapshots, despite minimal sampling rates, 4-bit quantization, and short durations (200\,ms–2\,s), through a hybrid coherent/non-coherent acquisition stage compensating for code Doppler.
These observations are exploited to tune simulation tools and to perform extended simulation campaigns,  
showing that the inclusion of additional constellations significantly improves availability; for a 26 dB-Hz acquisition threshold, the fraction of epochs with at least four visible satellites increases from 11\% to 46\% of the total epoch count.
These findings indicate that BeiDou, RNSS, and SBAS signals can substantially enhance GNSS-based autonomy for lunar and cis-lunar missions.
\end{abstract}




\section{Introduction}
As mission designs grow more complex and the number of planned lunar missions increases~\cite{spacenews_novaspace_2025}, the demand for navigation solutions that provide greater onboard autonomy in orbit determination and can scale with future mission needs has become a practical and strategic necessity. Historically, spacecraft operating beyond Earth orbit have depended primarily on ground-based \gls{RF} tracking networks, such as \gls{DSN}, \gls{ESTRACK}, \gls{ISTRAC}, and \gls{CDSN} to support navigation, guidance, and maneuvering. In parallel, recent progress in lunar exploration has highlighted the potential of \gls{GNSS} as a complementary resource for navigation in the lunar domain, encompassing cis-lunar space and the lunar surface, making the exploitation of these signals an increasingly compelling complement \cite{JournalDiMissione}.

In particular, the \gls{LuGRE} \cite{parker2022lunar, konitzer2024science} --- a joint NASA–ASI experiment hosted on Firefly Aerospace Inc.'s \gls{BGM1} lunar lander under NASA’s \gls{CLPS} program ---  was designed to acquire and track \gls{GPS} and Galileo signals in the L1/E1 and L5/E5a frequency bands along its trajectory to the Moon and during lunar surface operations, thereby exploiting the resulting observables to estimate the onboard receiver state, i.e., computing its \gls{PVT}. Launched in January 2025, \gls{LuGRE} achieved historic milestones, including multiple GNSS-only \gls{PVT} fixes along its trajectory and the first GNSS-only position fix on the lunar surface~\cite{JournalDiMissione}.

Beyond computing onboard \gls{PVT} solutions, the \gls{LuGRE} payload \cite{tedesco2023deep, pulliero2023space} collected more than 106~hours of unprecedented raw GNSS observables, as well as approximately $12$\,s of \gls{IQS} snapshots, with snapshot durations ranging from 200\,ms to 2\,s, collected across different operational windows. As a result, the openly available LuGRE dataset \cite{parker2025lunar} provides, in addition to state-estimation solutions, both \gls{IQS} snapshots and \gls{GNSS} receiver observables collected throughout the mission.

This paper leverages the unique \gls{IQS} snapshots obtained by \gls{LuGRE} \cite{parker2022lunar, konitzer2024science}, to demonstrate the availability of additional navigation signals from systems not originally exploited during mission operations, i.e., BeiDou, the \gls{QZSS}, the \gls{IRNSS} (also known as NavIC) and \gls{SBAS} constellations. A list of the constellations transmitting \gls{SBAS} signals is reported in Table \ref{tab:SBASlist}.

Beyond the experimental demonstration itself, the information extracted from the openly available \gls{IQS} snapshots \cite{parker2025lunar} is used to calibrate and validate an advanced extended \gls{SSV} simulator. This calibration, grounded directly on in-space experimental data, enables realistic simulation of \gls{LuGRE} operations while explicitly accounting for the contribution of additional \gls{GNSS}, \gls{RNSS}, and \gls{SBAS} constellations not processed during \gls{LuGRE} operations. As a result, the simulator provides, for the first time, an experimentally anchored assessment of multi-constellation signal availability in the cis-lunar and lunar environments.

The remainder of the article is organized as follows. Section \ref{sec:Background} describes the relevant features of the LuGRE payload and \gls{SC} operations. Section \ref{sec:acqStage} describes the acquisition stage utilized for processing snapshots of \gls{IQS} and the extended \gls{SSV} simulator employed to assess the experimental outcomes. Section \ref{sec:ResMain} analyzes acquisition results and multi-frequency, multi-constellation signal availability, while Section \ref{sec:conclusion} provides the conclusions.

\begin{table}[t]
\centering
\caption{List of constellations transmitting \gls{SBAS} signals (source: \gls{IGS}).}
\label{tab:SBASlist}
\begin{tabular}{lc}
\toprule
\textbf{System} & \textbf{Number of satellites} \\
\midrule
Al-SBAS  & 1 \\
BDSBAS   & 3 \\
EGNOS    & 3 \\
GAGAN    & 3 \\
KASS     & 2 \\
MSAS     & 2 \\
NSAS     & 1 \\
Pak-SBAS & 1 \\
QZSS     & 5 \\
SouthPAN & 1 \\
SDCM     & 3 \\
WAAS     & 3 \\
\bottomrule
\end{tabular}
\end{table}

\section{Background}
\label{sec:Background}

Several experiments have demonstrated the feasibility of using \gls{GNSS} signals beyond \gls{LEO}. Notably, \gls{NASA}'s \gls{MMS} mission exploited \gls{GPS} signals at distances of up to approximately 50\% of the Earth–Moon distance~\cite{mms2016navi,winternitz2017new,mmsRecordNews}. Complementarily, the \gls{CanX-2} mission demonstrated \gls{SBAS}-based ranging in \gls{LEO} \cite{kahr2016analysis, sarda2010canadian}, revealing the need for dedicated corrections to mitigate residual clock offsets on the order of 100\,m and for ephemerides more accurate than those provided by the \gls{SBAS} \glspl{SV}.
The missions discussed above did not lead to the release of publicly available datasets in the form of signal samples. The \gls{LuGRE} mission, instead, provides openly accessible data \cite{parker2025lunar}. Accordingly, this section presents an overview of \gls{LuGRE}, with particular emphasis on the \gls{LuGRE} payload and the \gls{IQS} data collection operations.

\subsection{The LuGRE Payload}
The \gls{LuGRE} payload consisted of the Qascom QN400-SPACE (QN400-S) receiver, together with a dedicated \gls{HGA}, a \gls{LNA}, and the harnesses required to connect all components. The receiver includes two independent boards controlled by a supervisory module to ensure complete redundancy. For most of the mission, the primary board handled all operations, while the secondary board was used only during the final two operations.

The payload’s \gls{HGA} was specifically designed for the reception of \gls{GNSS} signals in the L1/E1 and L5/E5a bands. It provides a peak boresight gain of 15.35\,dBic and includes an integrated notch filter to suppress out-of-band interference. During the transit phase operations, the antenna, mounted on a gimballed platform, was kept pointed toward the Earth by adjusting the spacecraft attitude while keeping the gimbal fixed, resulting in an Earth-Off-Pointing angle consistently below the nominal limit of 1°. During the surface phase, the active gimbal steering further improved alignment, achieving Earth-Off-Pointing angles better than 0.21° \cite{JournalDiMissione}. Consequently, antenna pointing error can be neglected for the analysis conducted in this work.

The \gls{LuGRE} payload was integrated into the \gls{BGM1} lander as a scientific instrument and operated in full compliance with strict do-no-harm requirements \cite{JournalDiMissione, pulliero2023space}. The lander did not rely on the payload for any spacecraft maneuvers, as the receiver was not interfaced with the \gls{GNC} system.

The QN400-S receiver supports a \gls{RTP} mode in which it autonomously handles signal acquisition, tracking, and state estimation using \gls{GPS} and Galileo signals in the L1/E1 and L5/E5a bands. At scheduled intervals, the receiver was configured in \gls{SC} mode, during which the receiver's front end collected raw \gls{IQS} snapshots on the same bands for subsequent downlink.

The receiver provided multiple \gls{ADC} configurations to serve for \gls{SC} mode. During standard \gls{RTP} operations, sampling was performed at 24\,Msps with 16-bit quantization. For L1/E1 acquisition, samples were decimated to 6\,Msps (16-bit), while the L5/E5a acquisition uses the full 24 Msps (16-bit). During tracking, L1/E1 signals are processed at 12\,Msps (16-bit), while L5/E5a continues at 24\,Msps (16-bit).

The \gls{IQS} snapshots collected in \gls{SC} mode were sampled at frequencies between 4\,MHz and 24\,MHz, with 4-bit or 8-bit quantization. Due to the limited downlink bandwidth shared among all payloads and subsystems, sampling and quantization choices directly impacted the available snapshot duration, limiting each snapshot to a maximum of two seconds, and higher-fidelity sampling configurations could not be used~\cite{nardin2023tracking}. Although the \gls{SC} and \gls{RTP} modes were mutually exclusive, the two modes were frequently executed in succession within the same operational window. Detailed timing, sequencing, and duration of these mode transitions for each operation are documented in the \gls{LuGRE} operations summary table (OPTABLE) contained in the mission’s public dataset \cite{parker2025lunar} and in~\cite{JournalDiMissione}.

\subsection{LuGRE Sample Capture Operations}
The start time of each \gls{SC} operation was selected to maximize the availability of \gls{GPS} and Galileo signals, when possible. Satellite visibility forecasts used during mission planning were generated using the \gls{GEONS} and \gls{GGMS} tools \cite{JournalDiMissione}.

The present article examines only data originated by operations featuring a \gls{SC}, and all results are obtained from the post-processing of the corresponding \gls{IQS} snapshots. Table~\ref{tab:OPTable} summarizes these operations, indicating, for each one, the \gls{ADC} configuration and the duration of the acquired signal snapshots. When both L1/E1 and L5/E5a \gls{IQS} snapshots were collected, the reported duration refers to both bands. Only operations in which the receiver was configured in \gls{SC} mode are listed.
The table also provides the \gls{UTC} start time of each \gls{SC} operation, denoted as $SC\ start$, together with the associated Julian day ($DoY$). \gls{UTC} timestamps assigned by the onboard computer to \gls{SC} operations, along with additional receiver status information, are available in the publicly released dataset \cite{parker2025lunar}.

\begin{figure}
    \centering
    \includegraphics[width=0.99\linewidth]{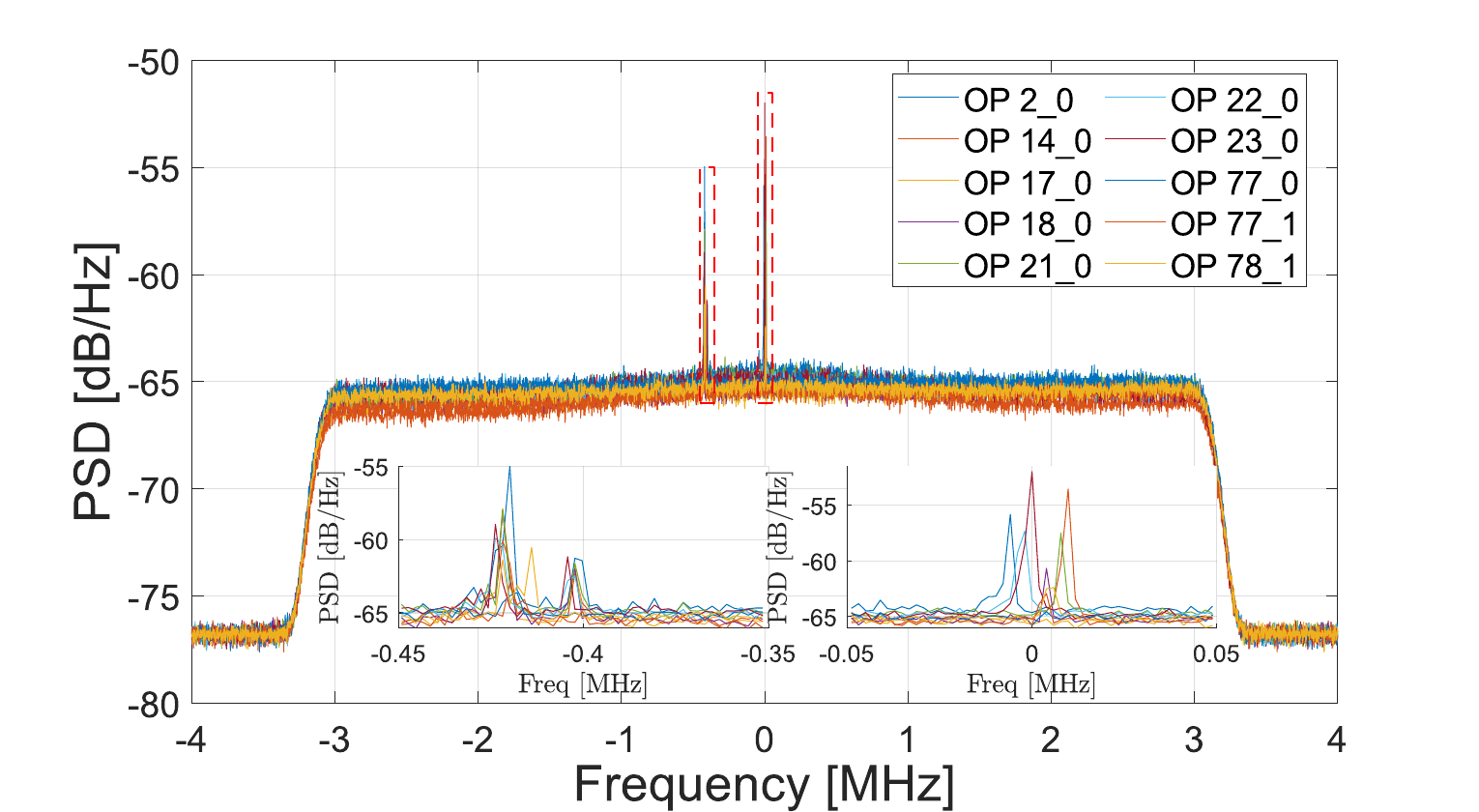}
    \caption{Example of \gls{PSD} obtained from the analysis of a subset of the \gls{IQS} snapshots. Quantization: 4 bits; sampling frequency: 8 MHz; band: L1.}
    \label{fig:IQ_PSD}
\end{figure}

\begin{figure}
    \centering
    \includegraphics[width=0.99\linewidth]{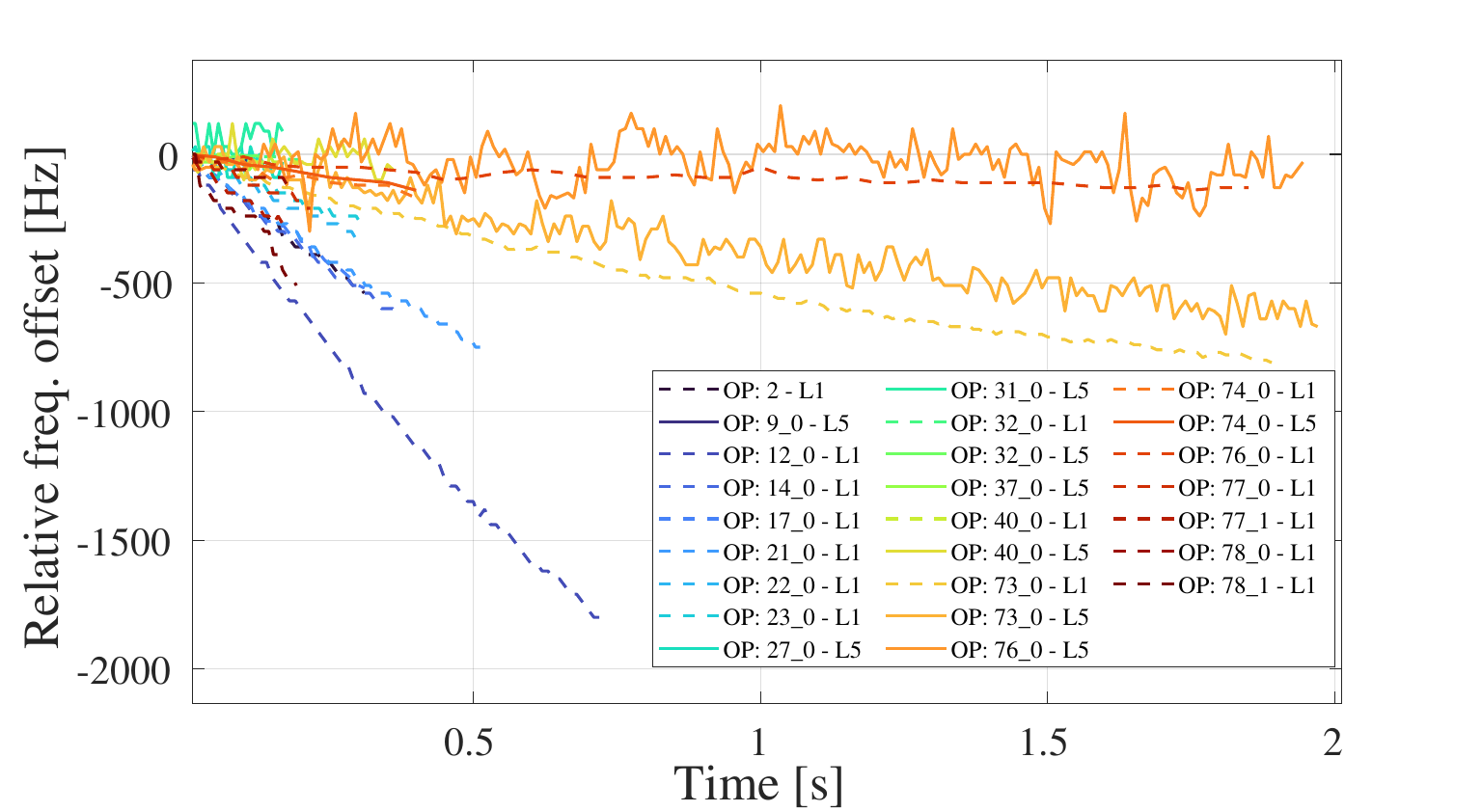}
    \caption{Frequency offset drift, expressed relatively to the initial frequency offset, observed over the  duration of each \gls{IQS} snapshot. Only a subset of the collected snapshots is included.}
    \label{fig:drift}
\end{figure}

Fig.~\ref{fig:IQ_PSD} shows an example of the estimated \gls{PSD} of the \gls{IQS} snapshots for one of the possible \gls{ADC} configurations on L1 band. As shown in \cite{JournalDiMissione}, the spurious frequencies observed are not expected to affect the measurement quality discussed in this work. Nonetheless, all operations exhibit a significant frequency drift that is inconsistent with the expected receiver dynamics.
For the sake of completeness, Fig.~\ref{fig:drift} presents the frequency drift relative to the start time of the \gls{SC} operation for a representative set of cases, where the frequency offset is estimated over multiple \gls{IQS} snapshot windows. 
The frequency offset drift is expected to affect the acquisition stage inducing a reduced performance by introducing a correlation loss dependent on the acquisition stage parameters. However, the \gls{RTP} dataset is less affected by the drift \cite{JournalDiMissione}, it is likely that the drift is introduced by the receiver clock when a reset of the receiver is performed, this happens when selecting the \gls{SC} and at the \gls{RTP} mode as well; anyway in \gls{RTP} mode the receiver clock requires many seconds between the acquisition and the beginning of the signal tracking to stabilize itself.
A further analysis on the root causes is out of scope and left for future works.

\begin{table*}[t]
\small
\caption{Summary of LuGRE's \gls{SC} operations.}
\label{tab:OPTable}
\centering
\begin{tabular}{l p{1.5cm} c p{2.5cm} p{2cm} l}
\toprule
\textbf{Mission phase} & \textbf{OP\newline identifier} & \textbf{DoY\newline (2025)} & \textbf{SC start\newline (UTC hh:mm:ss)} & \textbf{Snapshot\newline duration [ms]} &
\textbf{SC configuration} \\
\midrule
\multirow[t]{2}{*}{Commissioning}
 & 2\_0  & 16 & 01:31:11 & 400  & 4 bit, 8 MHz L1/E1 \\
\midrule
\multirow{7}{*}{Transit}
 & 5\_0  & 19 & 04:52:03 & 200  & 8 bit, 8 MHz L1/E1 \\
 & 9\_0  & 25 & 05:08:14 & 200  & 4 bit, 24 MHz L5/E5 \\
 & 12\_0 & 30 & 04:25:01 & 800  & 4 bit, 4 MHz L1/E1 \\
 & 14\_0 & 30 & 22:40:47 & 400  & 4 bit, 8 MHz L1/E1 \\
 & 17\_0 & 34 & 09:13:32 & 400  & 4 bit, 8 MHz L1/E1 \\
 & 18\_0 & 36 & 22:59:01 & 600  & 4 bit, 8 MHz L1/E1 \\
 & 21\_0 & 38 & 23:29:01 & 600  & 4 bit, 8 MHz L1/E1 \\
 & 22\_0 & 43 & 03:47:46 & 400  & 4 bit, 8 MHz L1/E1 \\
\midrule
\multirow{5}{*}{Lunar Orbit}
 & 23\_0 & 45 & 04:58:44 & 400  & 4 bit, 8 MHz L1/E1 \\
 & 27\_0 & 50 & 15:22:40 & 200  & 4 bit, 24 MHz L5/E5 \\
 & 31\_0 & 54 & 03:40:08 & 200  & 4 bit, 24 MHz L5/E5 \\
 & 32\_0 & 55 & 12:04:39 & 300  & 4 bit, 24 MHz L5/E5; 4 bit, 8 MHz L1/E1 \\
 & 37\_0 & 58 & 16:09:27 & 300  & 4 bit, 24 MHz L5/E5; 4 bit, 8 MHz L1/E1 \\
\midrule
\multirow{8}{*}{Lunar Surface}
 & 38\_0 & 62 & 06:12:50 & 300  & 4 bit, 24 MHz L5/E5; 4 bit, 8 MHz L1/E1 \\
 & 40\_0 & 63 & 07:03:12 & 400  & 4 bit, 24 MHz L5/E5; 4 bit, 8 MHz L1/E1 \\
 & 73\_0 & 73 & 10:09:35 & 2000 & 4 bit, 24 MHz L5/E5; 4 bit, 8 MHz L1/E1 \\
 & 74\_0 & 73 & 12:47:07 & 500  & 4 bit, 24 MHz L5/E5; 4 bit, 8 MHz L1/E1 \\
 & 76\_0 & 74 & 13:07:16 & 2000 & 4 bit, 24 MHz L5/E5; 4 bit, 8 MHz L1/E1 \\
 & 77\_0 & 75 & 15:12:22 & 300  & 4 bit, 8 MHz L1/E1 \\
 & 77\_1 & 75 & 19:14:55 & 300  & 4 bit, 8 MHz L1/E1 \\
 & 78\_0 & 75 & 22:03:52 & 300  & 4 bit, 8 MHz L1/E1 \\
 & 78\_1 & 75 & 22:11:20 & 300  & 4 bit, 8 MHz L1/E1 \\
\midrule
\bottomrule
\end{tabular}
\vspace{3pt}
\end{table*}

\section{Methodology}
\label{sec:acqStage}
\subsection{Signal Acquisition Architecture and Methods}
\label{sec:acqStrategyMeth}
\begin{figure}
    \centering
    \includegraphics[width=0.8\linewidth]{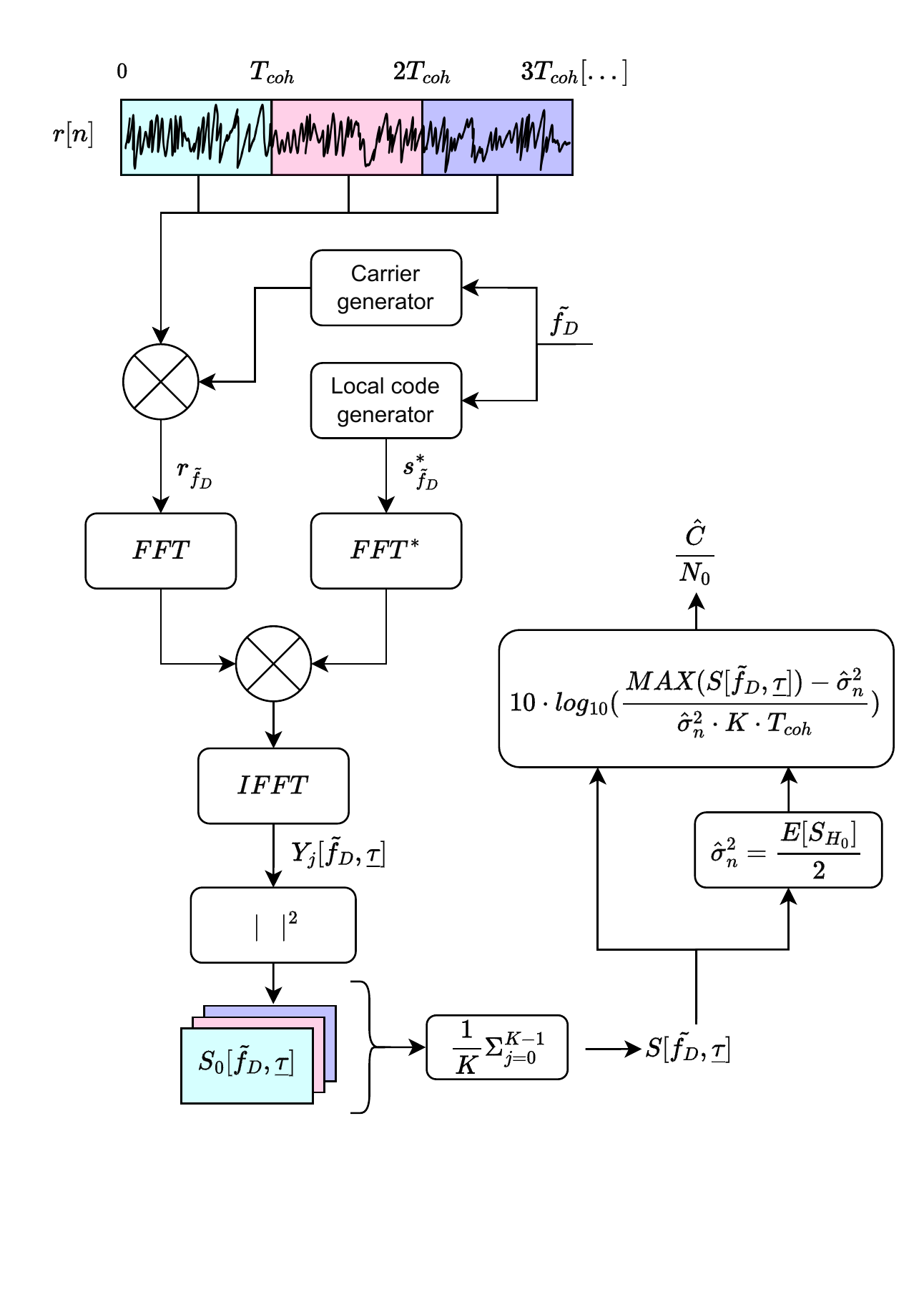}
    \caption{Schematic representation of the acquisition scheme.}
    \label{fig:AcqScheme}
\end{figure}

The adopted signal acquisition method is based on a hybrid coherent and non-coherent integration scheme with code Doppler compensation, as investigated in previous analyses~\cite{nardin2022acquisition}. For background on standard \gls{GNSS} signal acquisition architectures and methods, the interested reader is referred to~\cite{kaplan2017understanding}. 
The set of acquisition parameters chosen for the acquisition of the \gls{GNSS}, \gls{RNSS}, and \gls{SBAS} signals under analysis is provided in Table \ref{tab:paramAcqTot}. 
As reported in the literature \cite{van1991new, tsui2004fundamentals, borio2008statistical}, signal acquisition can be performed efficiently using a \gls{DFT}-based approach, which significantly reduces computational complexity. The mathematical formulation of the process is provided in the following, while a schematic representation of the \gls{CAF} computation and of the \gls{C/N0} estimation process is reported in Fig.~\ref{fig:AcqScheme}~\cite{kaplan2017understanding}.
Given the high dynamics of the scenario and the behavior of the payload clocks during operations, the Doppler effect and the  receiver clock drift significantly affect the correlation between the received and locally generated codes \cite{foucras2014detailed, sciacca2025acquiring}. As a result, the code Doppler must be accounted for when generating the local carrier and the local code.

To implement the \gls{DFT}-based acquisition scheme, the received signal $r[n]$ is multiplied by the local carrier testing for the frequency offset $\tilde f_D$ obtaining
\begin{equation}
    r_{\tilde{f}_D}[n] = r[n]\cdot e^{j2\pi(f_{IF}+ \tilde{f}_D)\frac{n}{f_s}}, \forall\ n\in\{0,..., K\cdot N-1\} ,
\end{equation}
where:
\begin{itemize}
    \item $f_{IF}$ is the intermediate frequency ($f_{IF}=0$ for \gls{LuGRE} \gls{SC} operations),
    \item $\tilde{f}_D$ is the frequency offset under test,
    \item $f_s$ is the sampling frequency, 
    \item $K$ is the number of non-coherent accumulation, 
    \item $N$ is the number of samples in a coherent integration time $T_{coh}=N/f_s$.
\end{itemize}

The acquisition metric $S[{\tilde f_D},\underline{\tau}]$, for the frequency offset under test $\tilde{f}_D$ is computed through a non-coherent integration of coherent correlation results, as detailed below. Specifically, the non-coherent integration is obtained as
\begin{equation}
S[{\tilde{f_D}},\underline{\tau}] = \frac{1}{K}\sum_{j=0}^{K-1} S_{j}[\tilde{f_D},\underline{\tau}],
\label{eq:fftacq3}
\end{equation}
where
\begin{equation}
S_{j}[\tilde{f_D},\underline{\tau}]= Y_{j}[\tilde{f_D},\underline{\tau}] \cdot Y_{j}^*[\tilde{f_D},\underline{\tau}] \
\label{eq:fftacq2}
\end{equation}
is the modulus-squared \gls{CAF} computed over $N$-sample signal sequences. $Y_{j}[\tilde{f_D},\underline{\tau}]$ is evaluated as
\begin{equation}
\begin{aligned}
Y_{j}[\tilde{f_D},\underline{\tau}]&= 
\mathrm{IDFT}\!\left\{
\mathrm{DFT}(r_{\tilde{f_D}}[jN, \ldots, jN+N-1]) 
\cdot \right.\\
&\qquad\left.
\mathrm{DFT}(s_{\tilde{f_D}}[jN, \ldots, jN+N-1])^*
\right\},
\end{aligned}
\label{eq:fftacq1}
\end{equation}
where
\begin{equation}
    \underline{\tau} = [0,...,N-1]\
\end{equation}
and where $DFT$ and $IDFT$ are respectively the discrete Fourier transform and the inverse discrete Fourier transform operators.
The local code $s_{\tilde f_D}$ is generated with a chip rate
\begin{equation}
    R_D = R_{chip} \left(1+\frac{\tilde f_D}{f_0}\right),
\end{equation}
which compensates for the Doppler effect on the PRN code.

\subsection{Acquisition threshold}
Following the evaluation of the \gls{CAF}, the acquisition is considered successful when the correlation peak exceeds a predefined threshold. By fixing a target system probability of false alarm $p_{fa}$, the probability of false alarm per cell can be expressed as
\begin{equation}
p_{fa,cell} = 1 - (1 - p_{fa})^{\frac{1}{L}},
\end{equation}
where $L$ denotes the size of the search space, i.e., the total number of frequency-offset and code-phase bins.
The threshold can then be set to
\begin{equation}
    Th = \hat{\sigma}^2_n\cdot \chi^{-2}(K,p_{fa,cell}),
\end{equation}
where $\chi^{-2}$ is the inverse chi-square function.
The term $\hat{\sigma}^2_n$ refers to the estimated noise variance measured at the acquisition stage.
To obtain $\hat{\sigma}^2_n$, according to the literature \cite{ betz2001effect, gomez2018c}, it is possible to consider the correlation under the null hypothesis $H_0$ in the Neyman-Pearson sense \cite{neyman1933ix} as the portion of \gls{CAF} $S_{H_0}$ located away from the correlation peak of at least one chip in the code phase domain. The expression of the noise variance estimate $\hat{\sigma}^2_n$ is
\begin{equation}
    \hat{\sigma}^2_n = \frac{E\left[ S_{H_0}\right]}{2}.
\end{equation}

For all operations, in this work, the system probability of false alarm is set to $0.001$. 
\begin{table}[h!]
\caption{Acquisition parameters used in post-processing during the acquisition campaign. Underlined values denote the settings adopted for the statistics reported in this work. Signals acquired with the secondary or overlay code are annotated accordingly.}
\label{tab:paramAcqTot}
\centering
\small
\begin{tabular}{l c c c}
\toprule
\textbf{Signal} & $\mathbf K$ \textbf{ non-coh. int.} & $\mathbf{T_{coh}\ (ms)}$ & \textbf{Secondary} \\
\midrule

GPS C/A        & \underline{15}, 20, 30                   & \underline{4}, 8            & --- \\
E1B            & \underline{15}, 20, 30, 40               & \underline{4}               & --- \\
B1C (Pilot)    & \underline{5}, 10, 15, 20                & \underline{10}              & \checkmark \\
QZSS L1 C/A    & \underline{15}, 20, 30                   & \underline{4}, 8            & --- \\
NavIC L1 (SPS) & \underline{5}, 10, 15, 20                & \underline{10}              & --- \\
L5Q            & \underline{20}                           & \underline{2}, 4, 8         & --- \\
E5aI           & \underline{20}                           & \underline{2}, 4, 8         & --- \\
B2aD (Data)    & \underline{10}, 15, 20, 30, 35           & \underline{5}               & \checkmark \\
QZSS L5I       & \underline{10}, 15, 20, 25               & \underline{10}              & \checkmark \\
NavIC L5       & \underline{20}                           & 4, \underline{8}            & --- \\
SBAS L1        & \underline{15}                           & \underline{4}               & --- \\
SBAS L5Q       & \underline{30}                           & \underline{1}               & --- \\
\bottomrule
\end{tabular}

\vspace{3pt}

\end{table}

\subsection{$C/N_0$ characterization}
Several methods for estimating \gls{C/N0} are available in the literature; some rely on the  outputs of the correlators at the tracking stage \cite{pini2008performance}, such as the \gls{MM} and \gls{NWPR} approaches. However, given the reduced length of the \gls{IQS} snapshots, it is relevant to reliably estimate \gls{C/N0} during the acquisition stage, for this purpose, the estimator chosen in this analysis is based on the \gls{NPDI} estimator introduced  in \cite{gomez2018c}.
The estimate of \gls{C/N0}, in dB-Hz, is obtained as
\begin{equation}
    \hat{\frac{{C}}{N_0}} = 10\cdot log_{10} \left( \frac{MAX(S[\tilde f_D, \underline{\tau}])-\hat\sigma^2_n}{\hat{\sigma^2_n}\cdot K\cdot T_{coh}}\right).
    \label{eq:CascoEstimator}
\end{equation}
This estimator is specifically intended for non-coherent integration and it is shown hereafter that it has a reduced bias in the range of \gls{C/N0} values considered in this work. 

\begin{figure}
\centerline{\includegraphics[width=18.5pc]{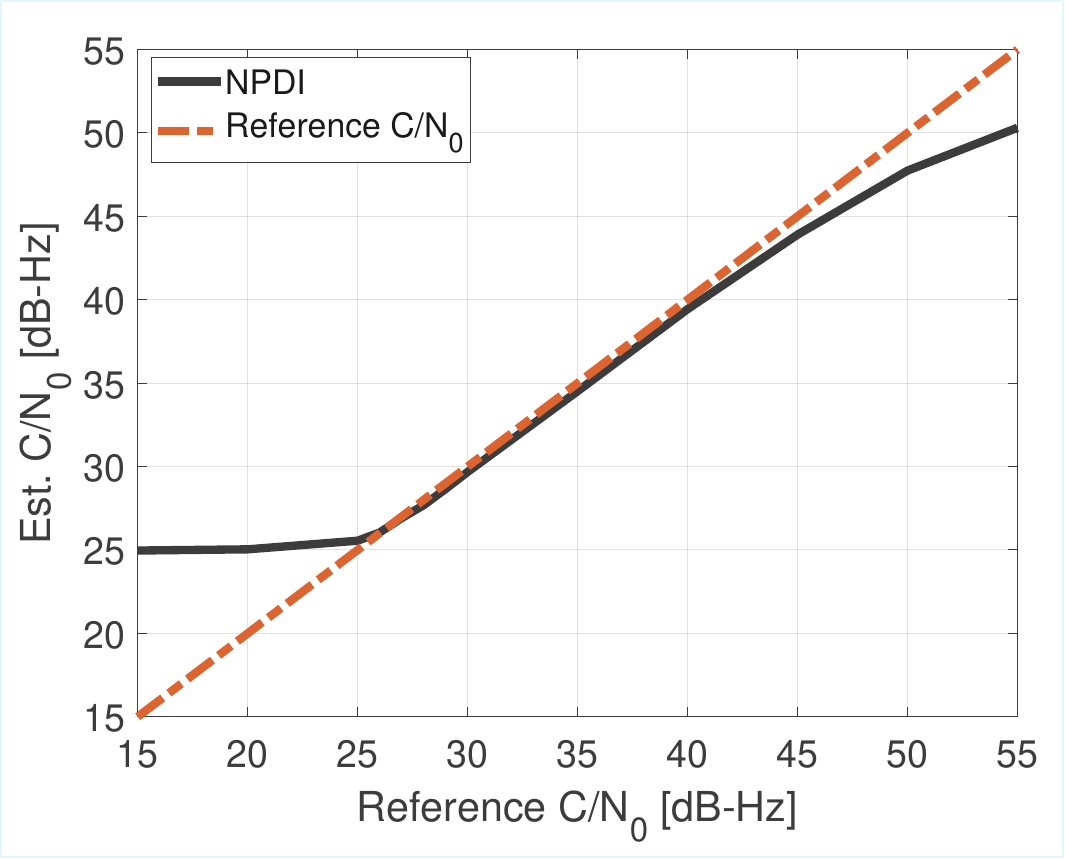}}
\caption{Comparison between expected and measured \gls{C/N0}. Generated \gls{C/N0} (orange), estimated \gls{C/N0} (black). Coherent integration time $T_{coh}:\ 4\ ms$, non coherent integrations: $K:\ 15$.}
\label{fig:CascoCompare}
\end{figure}

To understand how the bias introduced by the estimator affects the measurements, a set of simulations is conducted to evaluate the range in which the estimator is in its unbiased regime.
A synthetic GPS L1\,C/A signal is generated with a sampling frequency of $8$\,MHz and a quantization depth of 8\,bits. Fifteen \gls{C/N0} values, linearly spaced between 15 and 55\,dB-Hz, are tested. For each \gls{C/N0} level, 90 trials with independent noise realizations are evaluated using the estimator defined  in~\eqref{eq:CascoEstimator}. The number of non-coherent integrations is set to $K = 15$, with a coherent integration time $T_{\text{coh}} = 4\ \text{ms}$. 
As shown in Fig.~\ref{fig:CascoCompare}, the \gls{NPDI} based estimator exhibits a bias at higher simulated \gls{C/N0} values, causing the estimated value to negatively diverge from the true one. A similar behavior is also observed at lower \gls{C/N0} levels where estimated values become quickly unreliable as \gls{C/N0} drops below 25~dB-Hz. Taking into account the chosen acquisition parameters and the simulation setup, the estimated \gls{C/N0} range between 25 and approximately 40~dB-Hz are assumed reliable. Given the limited number of non-coherent integrations and the length of the coherent integration times adopted in this work, the \gls{C/N0} values that can be observed considering the receiver sensitivity are expected to fall within this range in most of the tested scenarios.  

To obtain an estimate of \gls{C/N0} from the available dataset, the estimation process is repeated for multiple adjacent and non-overlapping portions of the \gls{IQS} snapshots and then the results are averaged together to obtain a single \gls{C/N0} estimate.  

\subsection{Extended Space Service Volume Simulator}
The extended \gls{SSV} simulator used in this work models the \gls{GNSS}, \gls{RNSS}, and \gls{SBAS} constellations relevant to the analysis. It propagates broadcast ephemerides to determine satellite positions and generates the corresponding observables, including \gls{C/N0}, geometric Doppler shift, and pseudoranges. To estimate satellite visibility and received \gls{C/N0} for L1 C/A and L5 signals at the spacecraft location, the simulator assigns to each satellite an antenna radiation pattern derived from the available literature. Table~\ref{tab:patternTab} summarizes the pattern adopted for each constellation.

High-fidelity three-dimensional gain patterns are available for \gls{GPS}, Galileo, and \gls{QZSS} on both L1 and L5; these are directly incorporated into the simulation. For BeiDou, complete antenna patterns have not yet been published at the time of writing. The simulator therefore employs the Min-Model, which provides a conservative representation of the main lobe for each band \cite{unoosa_icg17_wgb_07} and reconstructs the corresponding pattern through interpolation of the published values.

For other \gls{RNSS} systems, detailed radiation patterns are similarly unavailable. In the case of \gls{GAGAN} and \gls{IRNSS}, an approximate antenna pattern is derived from CanX-2 mission measurements reported in \cite{kahr2016analysis}. This pattern is constructed by interpolating the available points conservatively, preserving the main lobe while assigning low gain at large off-boresight angles. For the remaining \gls{SBAS} constellations, except for \gls{GAGAN}, the BeiDou Min-Model is used for \gls{GEO} and \gls{IGSO} satellites.

To take full advantage of the LuGRE dataset, the link budget for each constellation is calibrated using measurements collected during the mission. For \gls{GPS} and Galileo, more than 106~hours of \gls{RTP} observations support a high-accuracy calibration. The procedure begins by assuming nominal transmit power values and running an initial simulation across all relevant operations. The residuals between the simulated and measured \gls{C/N0} values are evaluated considering only the epochs in which the signals from the main lobes of the transmitting satellites are in radiometric visibility. These residuals are then aggregated by \gls{PRN} and frequency band. The resulting biases, expressed as penalties or gains, are applied to the corresponding satellites to align the simulations with the observed data.

For the remaining \gls{GNSS}, \gls{RNSS}, and \gls{SBAS} constellations, calibration relies exclusively on the \gls{C/N0} estimates obtained from the \gls{IQS} snapshots collected during \gls{SC} operations. Although this provides a less accurate calibration, due to the limited number of available measurements and the absence of detailed antenna patterns, it enables, for the first time, a simulation of multi-constellation signal availability in the cis-lunar environment based directly on experimental observations acquired in cis-lunar space and at the lunar surface.
It is important to note that the power distribution among the various signal components for \gls{GPS}, as well as for other constellations, may vary over time, for instance, in the case of \gls{GPS}, due to the feature called flex power mode \cite{steigenberger2019flex}.

SBAS signals on L5 were excluded due to limited information and for the potentially large variety of L5 antenna patterns. Additionally, some \glspl{SV} transmit both ranging and augmentation signals, namely BeiDou GEO-1/2/3 and all \gls{QZSS} satellites except QZS-1. Their ranging signals count toward their constellations, while augmentation signals are treated separately and not counted as \gls{SBAS}.
It can be anticipated that the \glspl{PRN} acquired are associated to the signals that, according to the simulator outputs, are all coming from the mainlobes of the transmitting satellites and are all expected to be in radiometric visibility. 

\begin{table*}[h!]
\caption{Antenna patterns used in the extended \gls{SSV} simulator for each \gls{GNSS}, \gls{RNSS}, and \gls{SBAS} constellation. Patterns available per \gls{SV} on L1 or L5 bands are marked accordingly.}
\small
\begin{tabular}{l c p{6.5cm} p{3.5cm}}
\toprule
\textbf{Constellation} & \textbf{Per-SV Availability} & \textbf{Description} & \textbf{References} \\
\midrule

GPS & \checkmark$^*$ &
Official antenna pattern pre-flight measurements released by Lockheed Martin and Boeing. &
\cite{uscg_navcen_gps_tech_refs} \\
Galileo & --- &
GRAP model. &
\cite{menzione2024reconstruction} \\

BeiDou (BDS-3) & --- &
Min-model of the antenna main lobe, available per band and orbit type. &
\cite{unoosa_icg17_wgb_07} \cite{lin2020bds}\\

QZSS & \checkmark &
Official antenna pattern released by the \gls{JAXA}. &
\cite{qzss_antenna_patterns} \\

NavIC (IRNSS) & --- &
Pattern available only for a submodule of the antenna.  
The patterns used correspond to BeiDou BDS-3 (B1) GEO-IGSO. &
    \cite{RajeevGAGAN2005}\\

SBAS & --- & 
Patterns used in the simulations correspond to BeiDou BDS-3 (B1) GEO-IGSO **. &
--- \\
\bottomrule
\multicolumn{4}{l}{\footnotesize * Pattern available for each SV except block IIF (released at block level)  and for SVs with PRN 1, 11, 21, and 28, which are not available.} \\
\multicolumn{4}{l}{\footnotesize ** Except for \gls{GAGAN} \cite{kahr2016analysis}.} \\
\end{tabular}

\label{tab:patternTab}
\end{table*}

\section{Results and Analysis}
\label{sec:ResMain}

\subsection{Acquisition Results from IQ Snapshots Collected in the Lunar Domain}
\begin{figure*}[t]
    \centering
    \subfigure[L1 band\label{fig:AvailGPS_GAL}]{
        \includegraphics[width=0.48\textwidth]{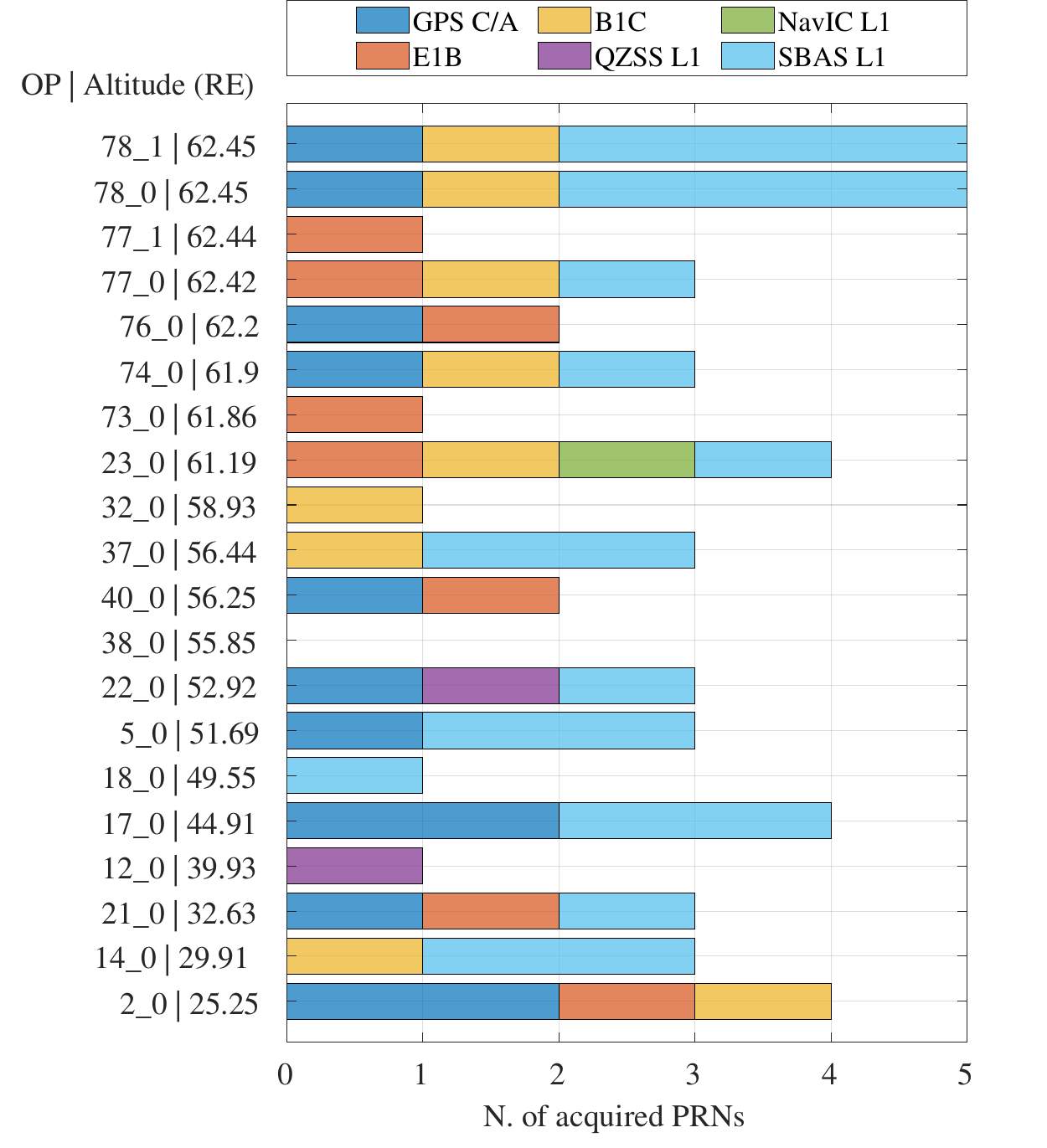}
       \label{fig:acqStack}
    }\hfill
    \subfigure[L5 band\label{fig:AvailRNSS_SBAS}]{
        \includegraphics[width=0.48\textwidth]{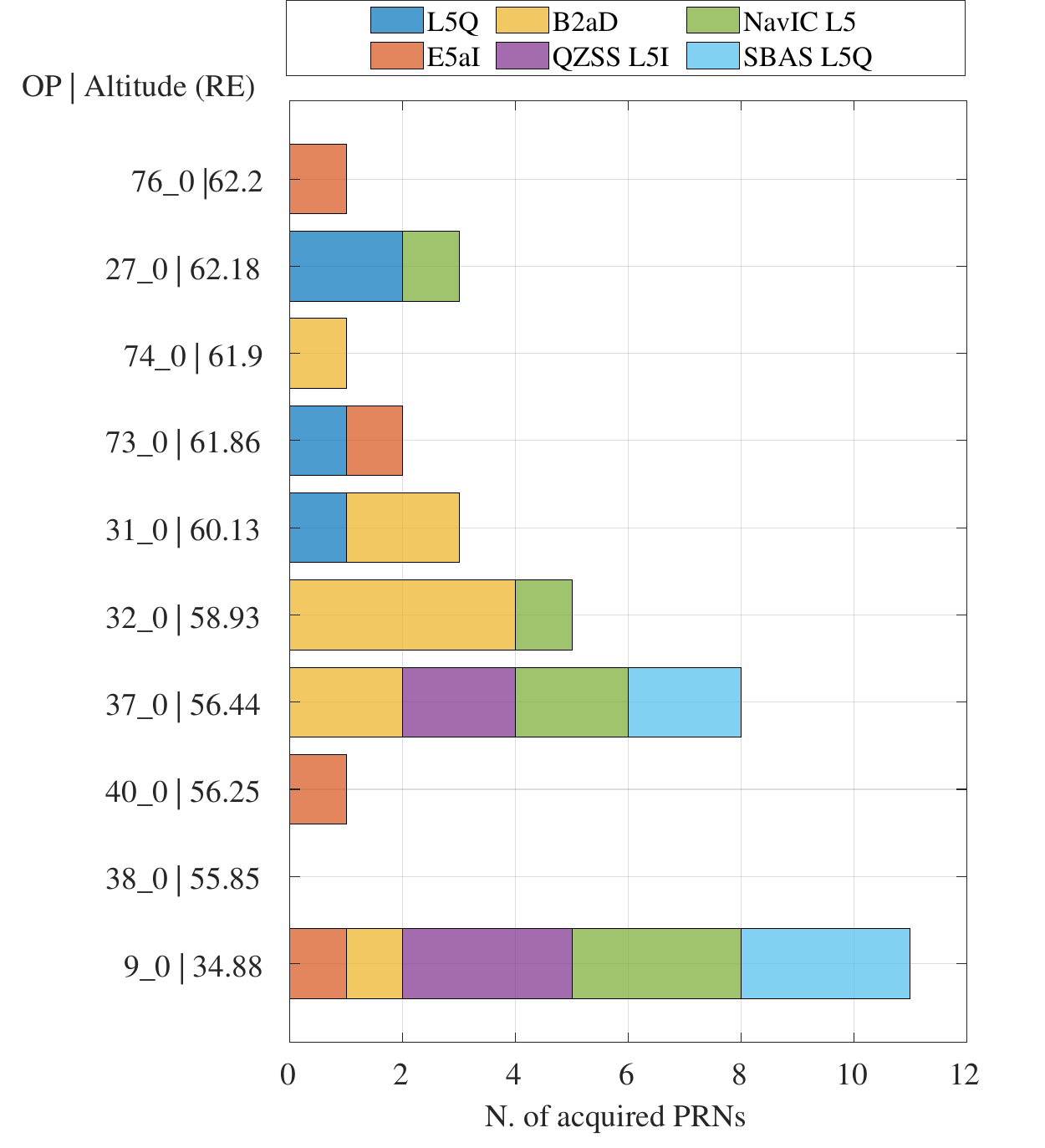}
           \label{fig:acqStackL5}
    }

    \caption{Number of acquired signals for each operation. Operations are ordered as function of altitude expressed in \gls{RE}.}
    \label{fig:Avail_multi}
\end{figure*}

A comprehensive overview of all successful acquisitions is reported in Table~\ref{tab:ACQRESULTS} and in Table~\ref{tab:ACQRESULTSL5} for L1/E1 and L5/E5a bands, respectively. Although the \gls{SC} operations were optimized solely for \gls{GPS} and Galileo visibility, a significant number of signals from the BeiDou \gls{RNSS} and various \glspl{SBAS} were acquired from the available \gls{IQS} snapshots. This result is illustrated in Fig.~\ref{fig:acqStack} and Fig.~\ref{fig:acqStackL5}, which show, for each \gls{SC} operation, the number of acquired \gls{PRN} signals per system and per band as a function of the spacecraft altitude at the start of the operation, expressed in \gls{RE}. Among all operations, OP38\_0 did not yield any successful acquisitions. In contrast, the snapshot associated with OP78\_1, despite being collected at a high altitude, produced the highest number of L1 acquisitions, while OP9\_0 resulted in the largest number of L5 acquisitions.

\begin{figure*}[t]
    \centering
    \subfigure[L1 band \label{fig:AvailGPS_GAL}]{
        \includegraphics[width=0.48\textwidth]{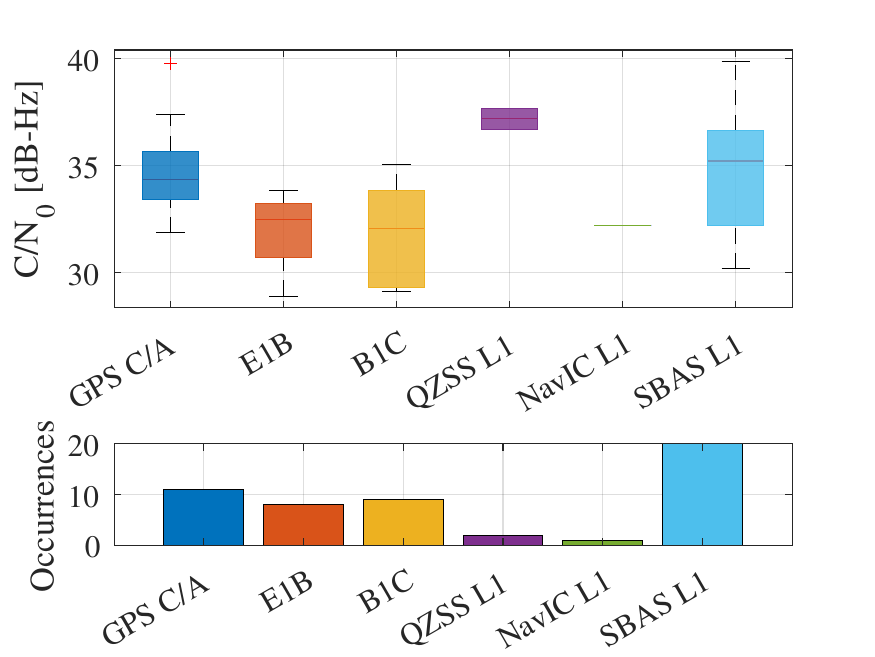}
         \label{fig:acqDistr1}
    }\hfill
    \subfigure[L5 band\label{fig:AvailRNSS_SBAS}]{
        \includegraphics[width=0.48\textwidth]{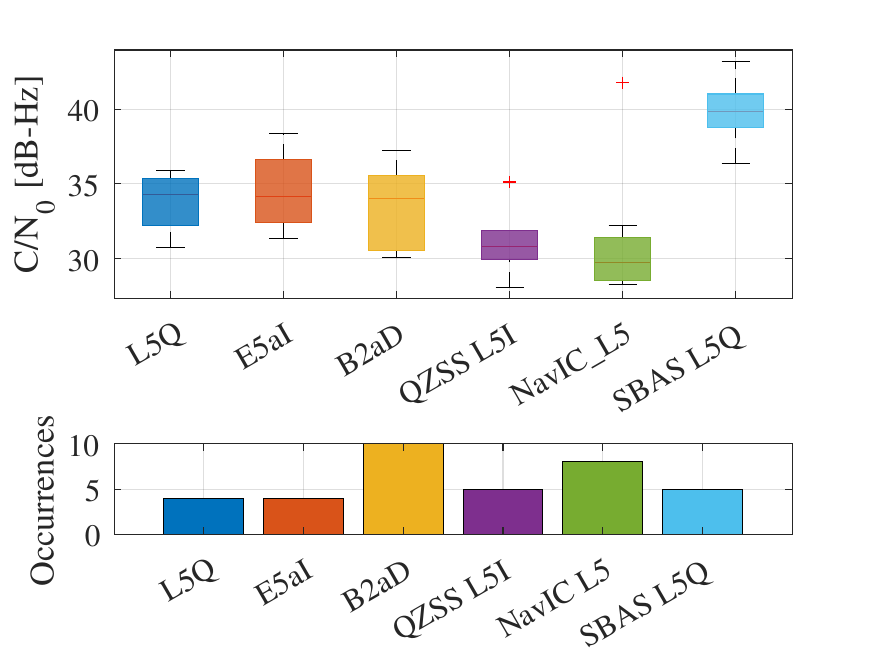}
         \label{fig:acqDistr5}
    }

    \caption{Distribution of the carrier-to-noise density ratio ($C/N_0$) for different GNSS signals in the L1  and L5 bands. For each signal, the central line of the box represents the median $C/N_0$, the box spans the interquartile range (25th--75th percentiles), and the whiskers extend to the most extreme values excluding outliers. Outliers, when present, are indicated by individual markers. The lower panel reports the number of occurrences for each signal.}
    \label{fig:Avail_multi}
\end{figure*}

Fig.~\ref{fig:acqDistr1} and Fig.~\ref{fig:acqDistr5} present the distribution of \gls{C/N0} values for each signal on the  L1 and L5 bands, together with the number of successful acquisitions per signal. The analysis indicates that expanding the set of processed signals to include those of additional \gls{GNSS}, \gls{RNSS}, and \gls{SBAS} constellations substantially increases the total number of successful acquisitions. On the L1 band, the largest contributions arise from \gls{SBAS} and BeiDou, whereas on the L5 band, BeiDou and \gls{IRNSS} are the most significant. Notably, the NavIC L5 signal is characterized by a chip rate of 1.023~Mchip/s, which is 10 times lower than the chip rate of all other L5 signals, making it uniquely oversampled. It is worth recalling that all L5 \gls{SC} operations were instead performed at a nominal sampling rate of 24~MHz.

\subsection{Simulated Signal Availability}
The acquisitions performed on the \gls{IQS} snapshots from \gls{LuGRE} were used to tune the extended \gls{SSV} simulator by adjusting the transmitted power of each constellation. This section presents the simulation results corresponding to the scenarios encountered by \gls{BGM1} during the \gls{LuGRE} operational windows.


For all subsequent analyses, a threshold of 26 dB-Hz is used. This value provides a conservative margin, since the \gls{LuGRE} receiver has successfully tracked signals with an estimated \gls{C/N0} of about 24 dB-Hz~\cite{JournalDiMissione}. The selected threshold also improves robustness against possible modeling inaccuracies and other sources of signal degradation. It is important to note that the results shown here, besides the selected sensitivity threshold, are receiver-agnostic; different receivers may adopt different acquisition and tracking strategies, and their effective thresholds can vary depending on both environmental conditions and internal receiver state.

Fig.~\ref{fig:AvailGPS_GAL} and Fig.~\ref{fig:AvailRNSS_SBAS} illustrate the percentage of the nominal RTP operation duration during which 
$n$ signals from $n$ distinct satellites exceed the selected \gls{C/N0} threshold. Specifically, Fig.~\ref{fig:AvailGPS_GAL} presents the multi-frequency, multi-constellation scenario including only \gls{GPS} and Galileo, while Fig.~\ref{fig:AvailRNSS_SBAS} shows the case where additional GNSS, \gls{RNSS}, and \gls{SBAS} constellations are also considered. The figures show an increase in the availability introduced by the additional constellations; however, some operations are still characterized by a reduced number of satellites in radiometric visibility, such as the case of~OP74\_0 and~OP27\_0. 
Including BeiDou, as well as the \gls{RNSS} and \gls{SBAS} constellations, increases the percentage of epochs with at least four satellites in radiometric visibility from 11.7\% to 46.1\%. The adopted threshold, i.e., $n\geq4$, was considered as it represents the minimum requirement for autonomous receiver state estimation, assuming the use of limited a priori information to guarantee solution convergence (e.g. through a Least-Squares estimation). This condition allows for the reliable initialization of more advanced orbit determination filters such as Bayesian estimation filters devoted to \gls{POD}. 

\subsection{Dilution of precision}
\begin{figure*}
    \centering
    \subfigure[OP2\_0]{\includegraphics[width=0.48\textwidth]{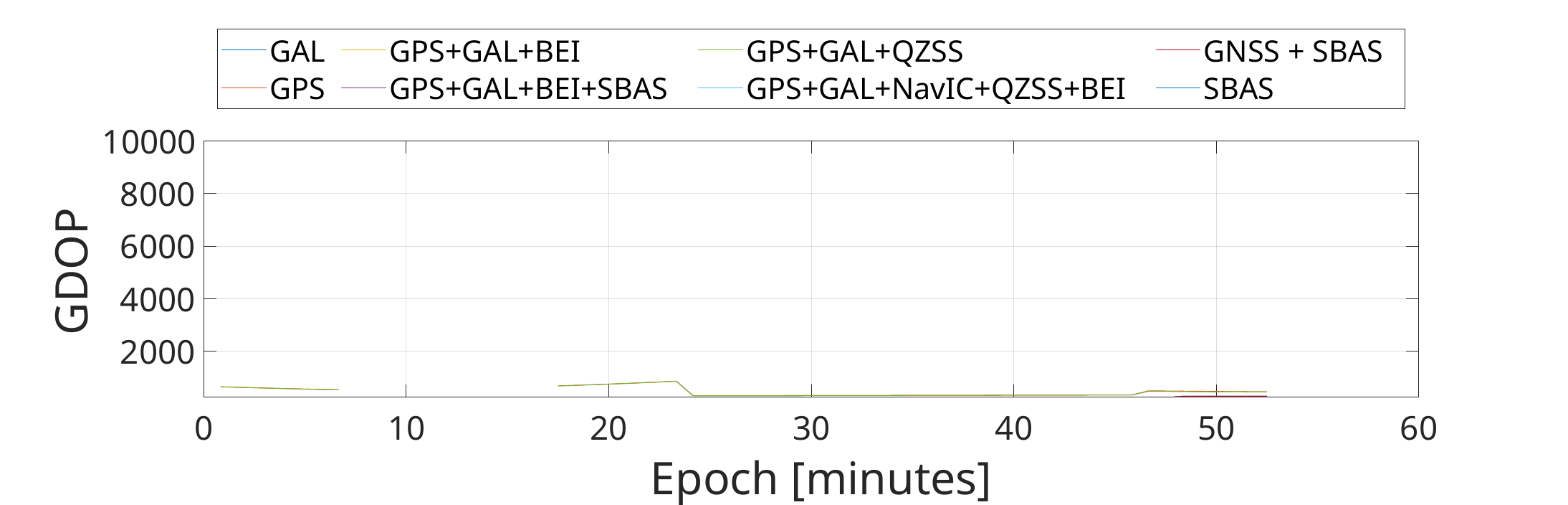}} 
    \subfigure[OP5\_0]{\includegraphics[width=0.48\textwidth]{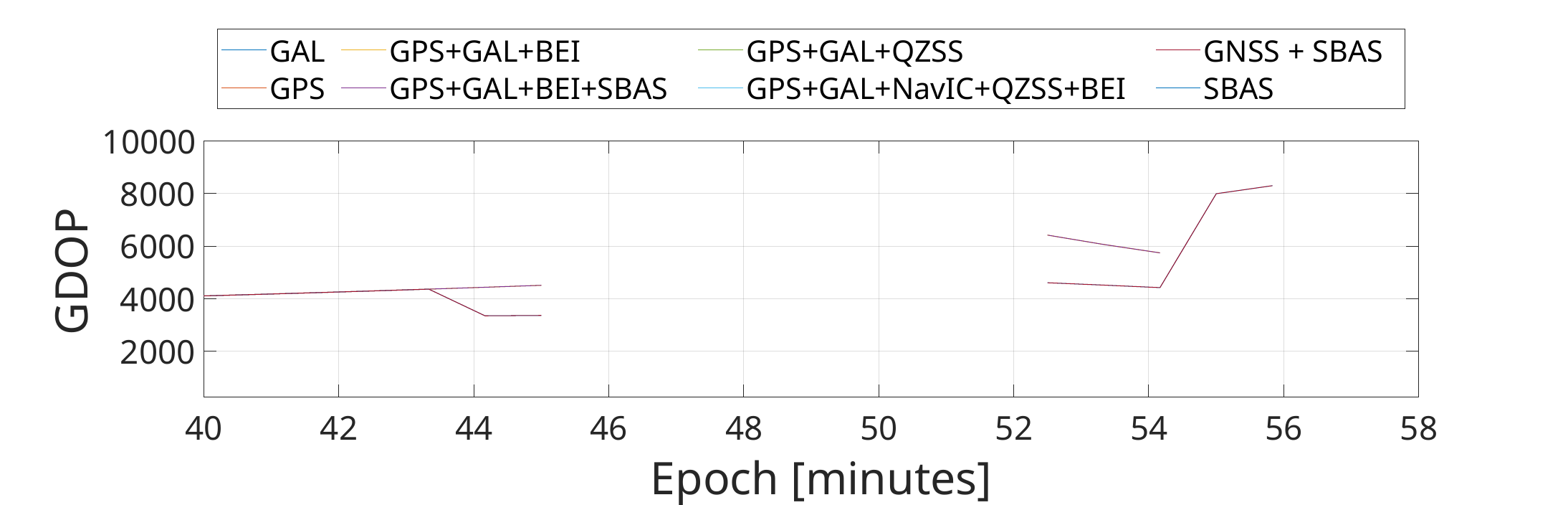}} 
    \subfigure[OP12\_0]{\includegraphics[width=0.48\textwidth]{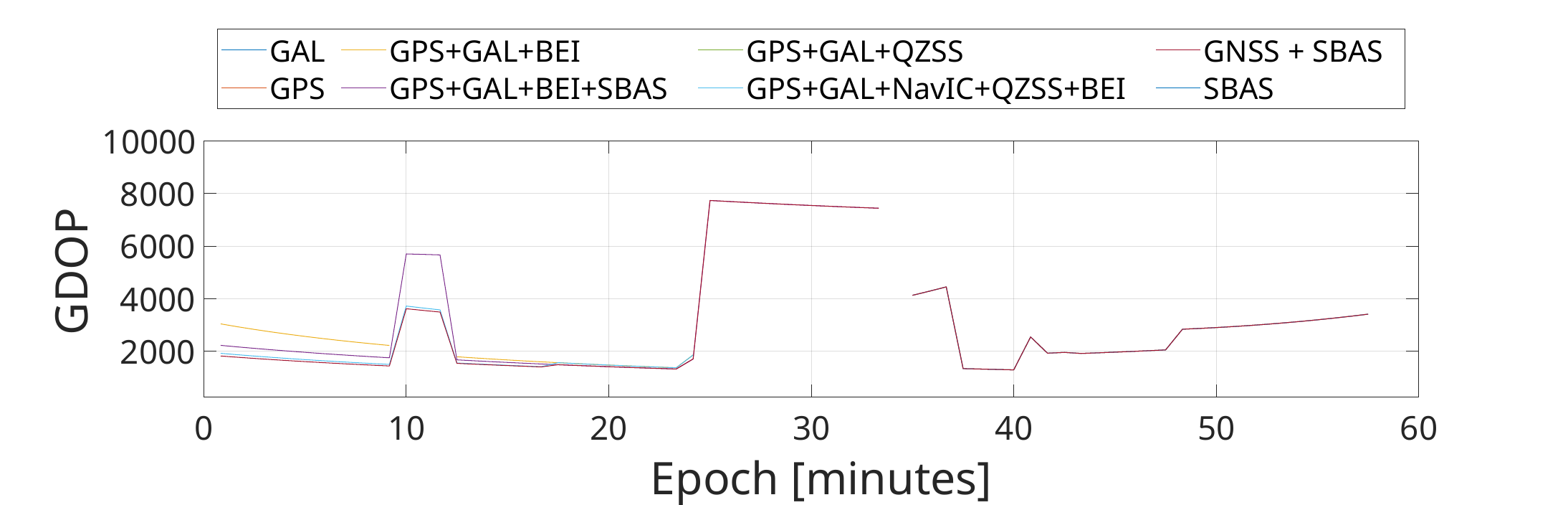}} 
        \subfigure[OP14\_0]{\includegraphics[width=0.48\textwidth]{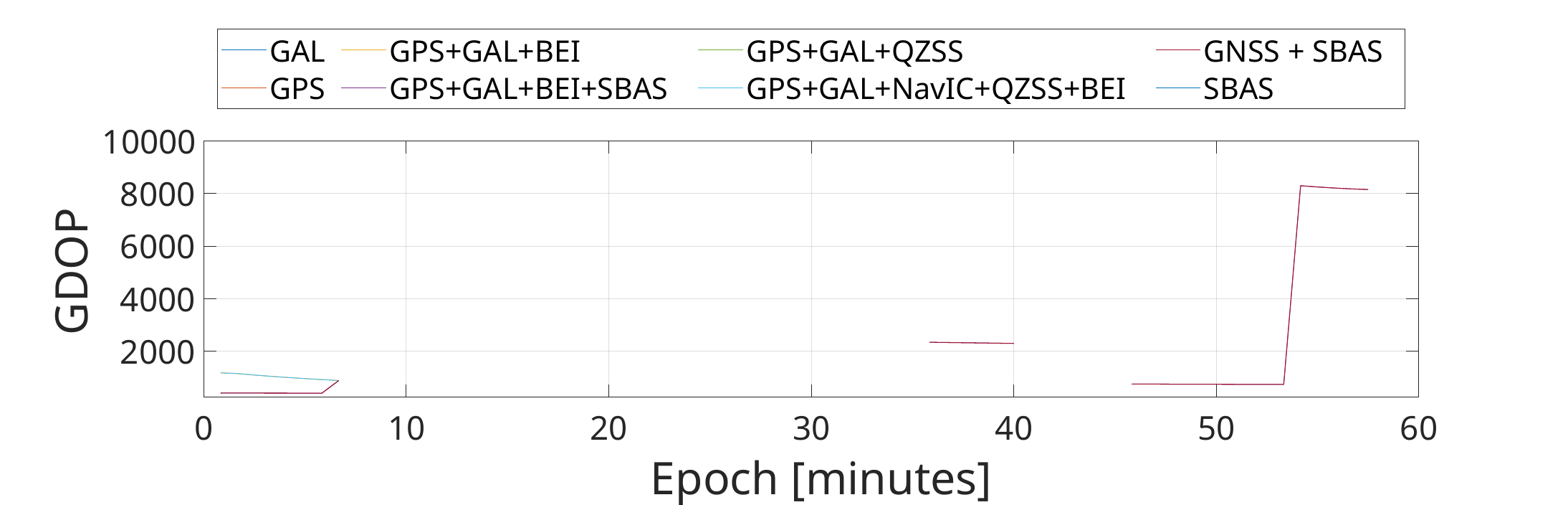}} 
    \subfigure[OP17\_0]{\includegraphics[width=0.48\textwidth]{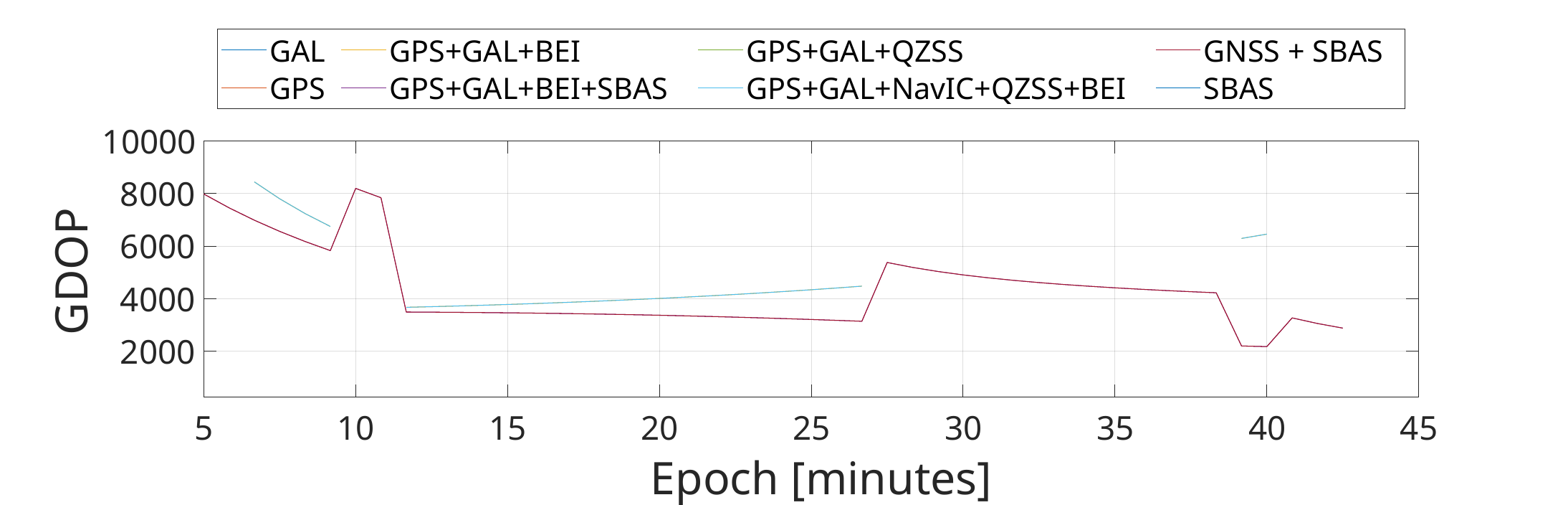}} 
     \subfigure[OP21\_0]{\includegraphics[width=0.48\textwidth]{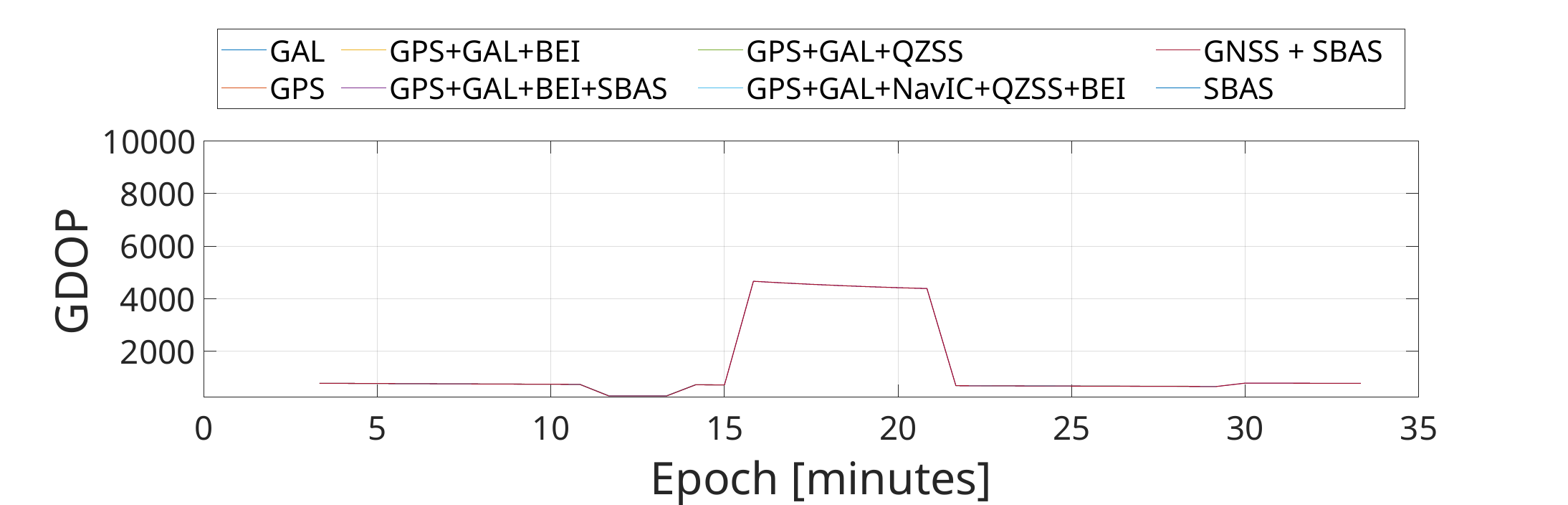}}
    \subfigure[OP22\_0]{\includegraphics[width=0.48\textwidth]{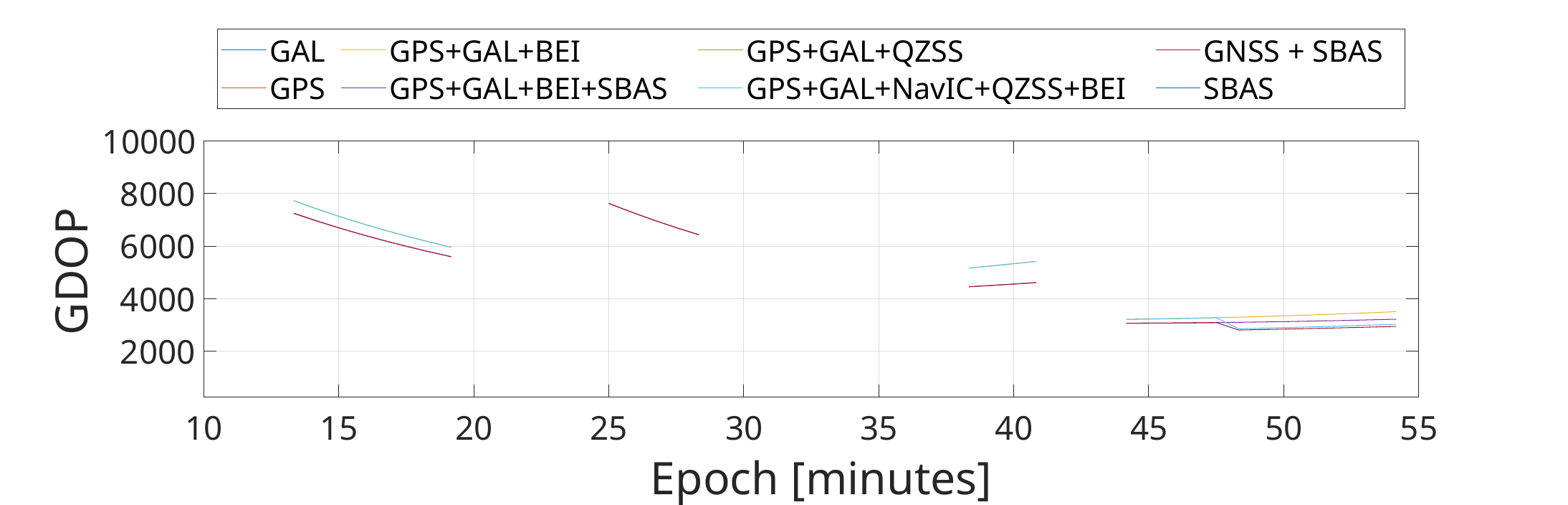}} 
    \subfigure[OP23\_0]{\includegraphics[width=0.48\textwidth]{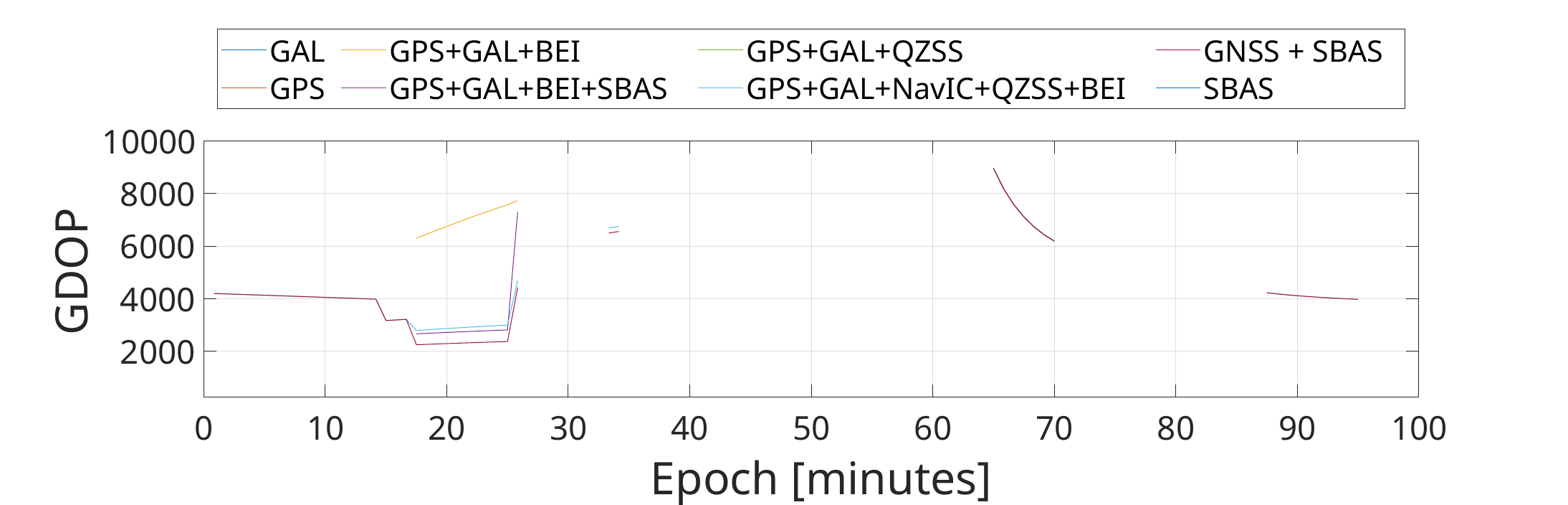}} 
    \subfigure[OP37\_0]{\includegraphics[width=0.48\textwidth]{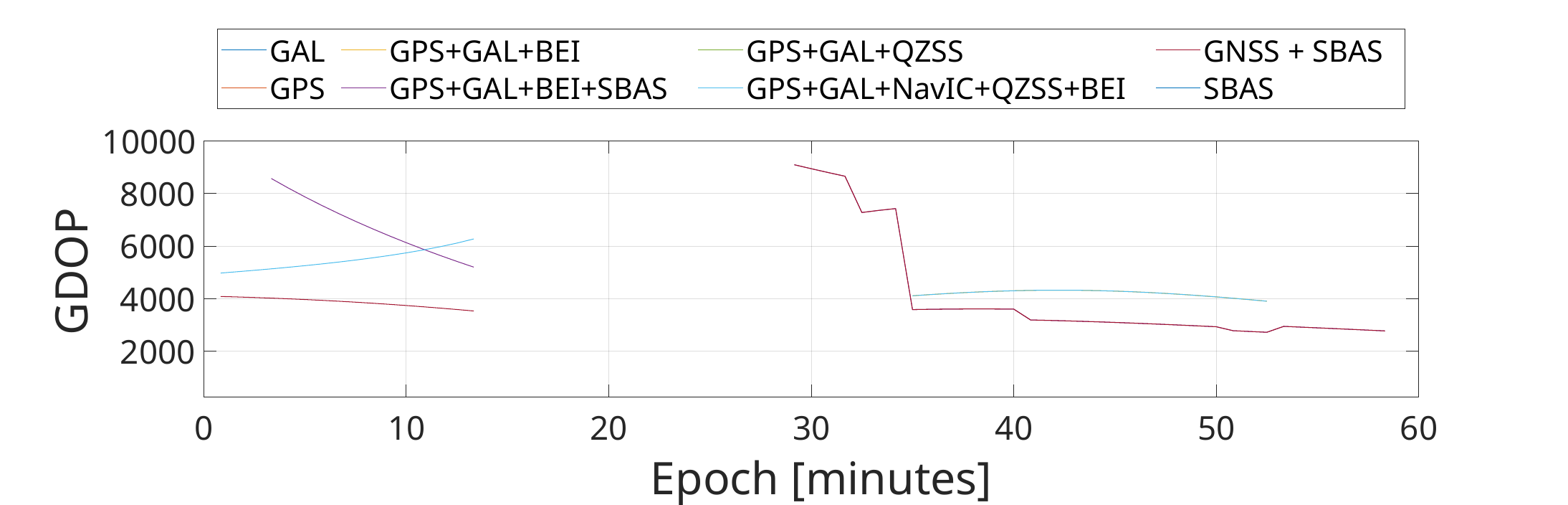}} 
    \subfigure[OP38\_0]{\includegraphics[width=0.48\textwidth]{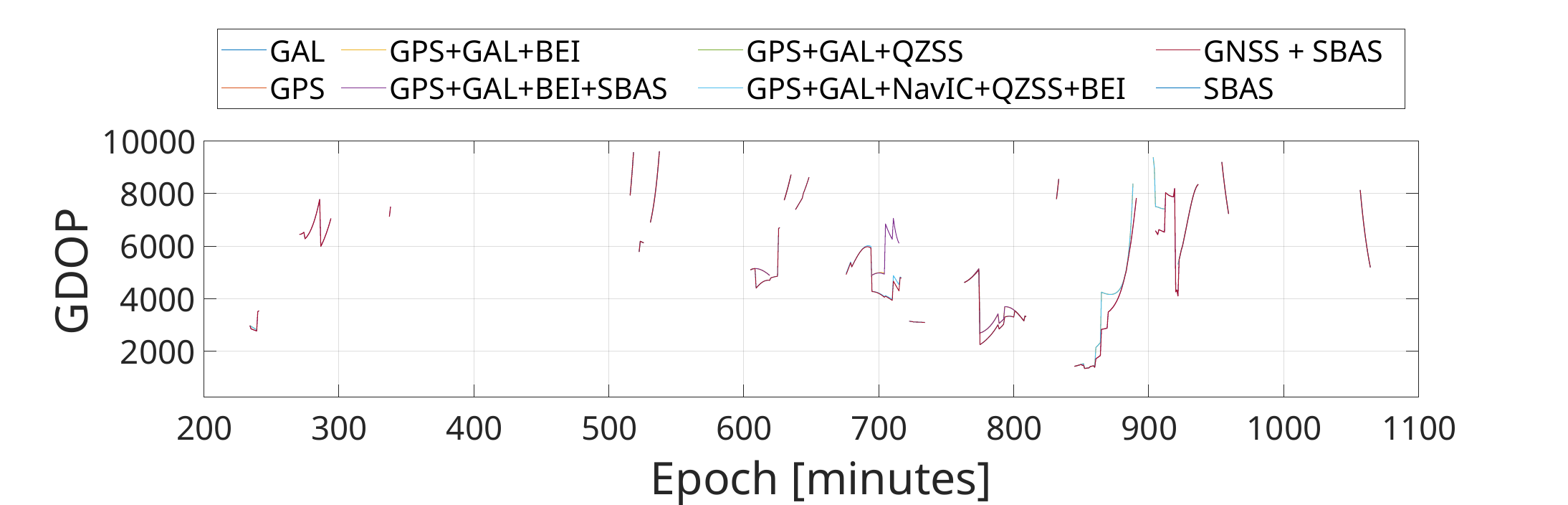}}
    \subfigure[OP40\_0]{\includegraphics[width=0.48\textwidth]{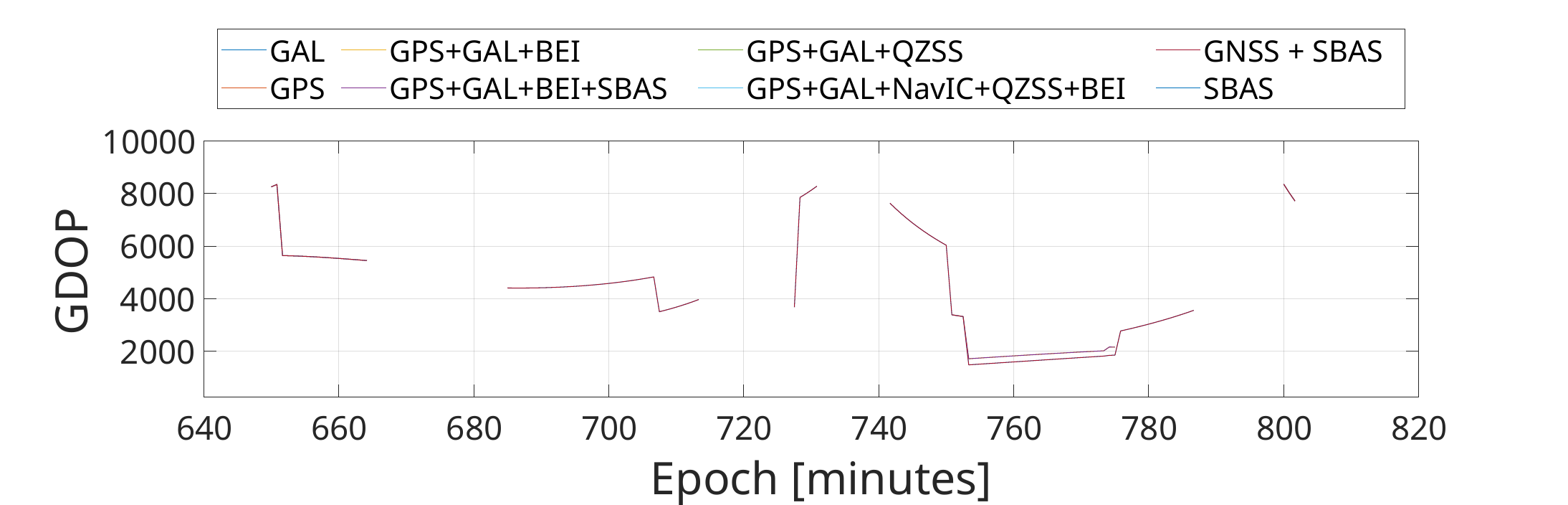}}
    \subfigure[OP76\_0]{\includegraphics[width=0.48\textwidth]{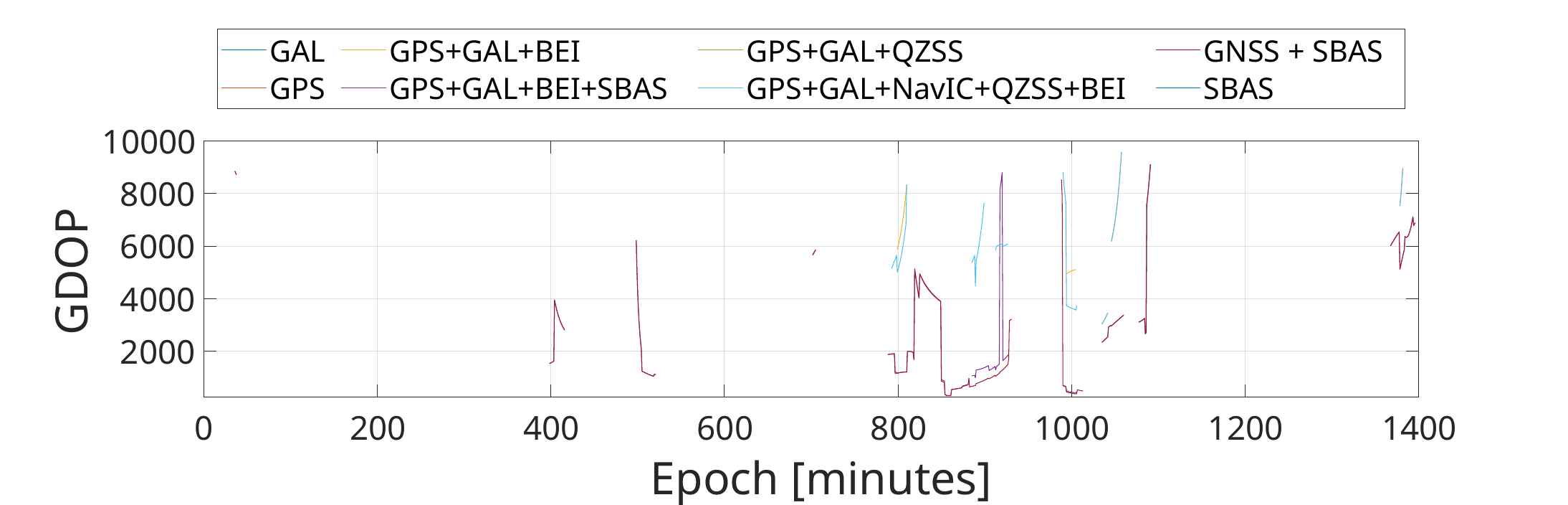}} 
    \subfigure[OP77\_1]{\includegraphics[width=0.48\textwidth]{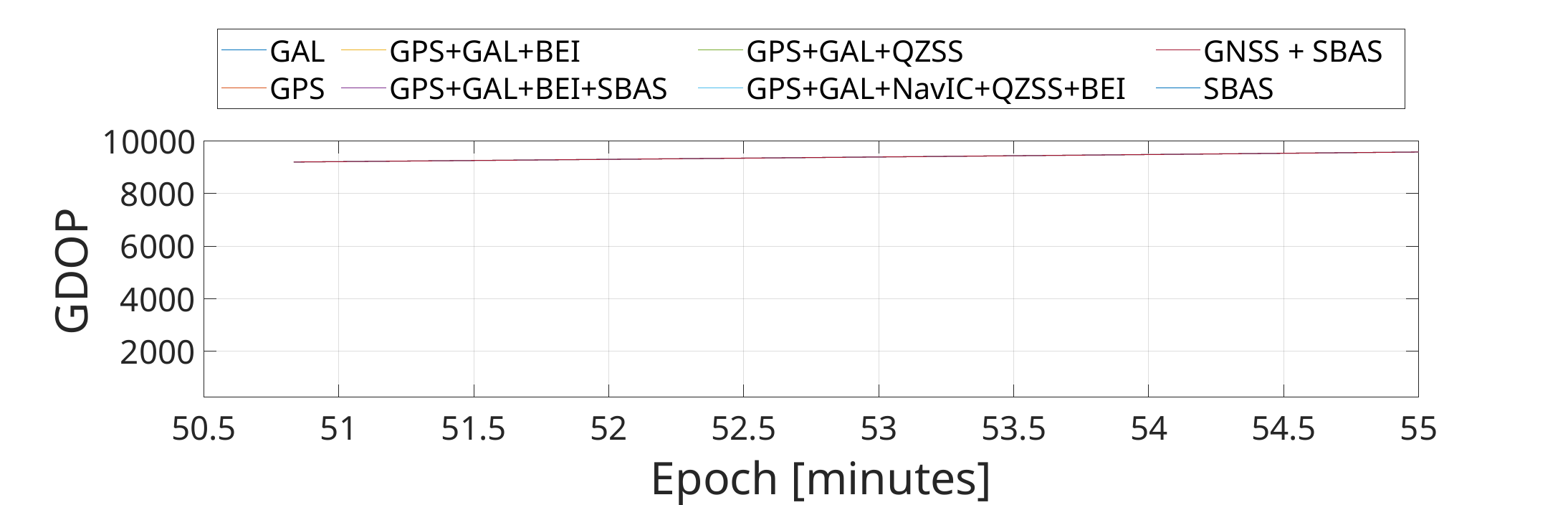}}      
    \caption{\gls{GDOP} values corresponding to the set of operations where evaluation was possible.}
    \label{fig:AllGDOP}
\end{figure*}

\begin{figure}
    \centering
    \includegraphics[width=0.98\linewidth]{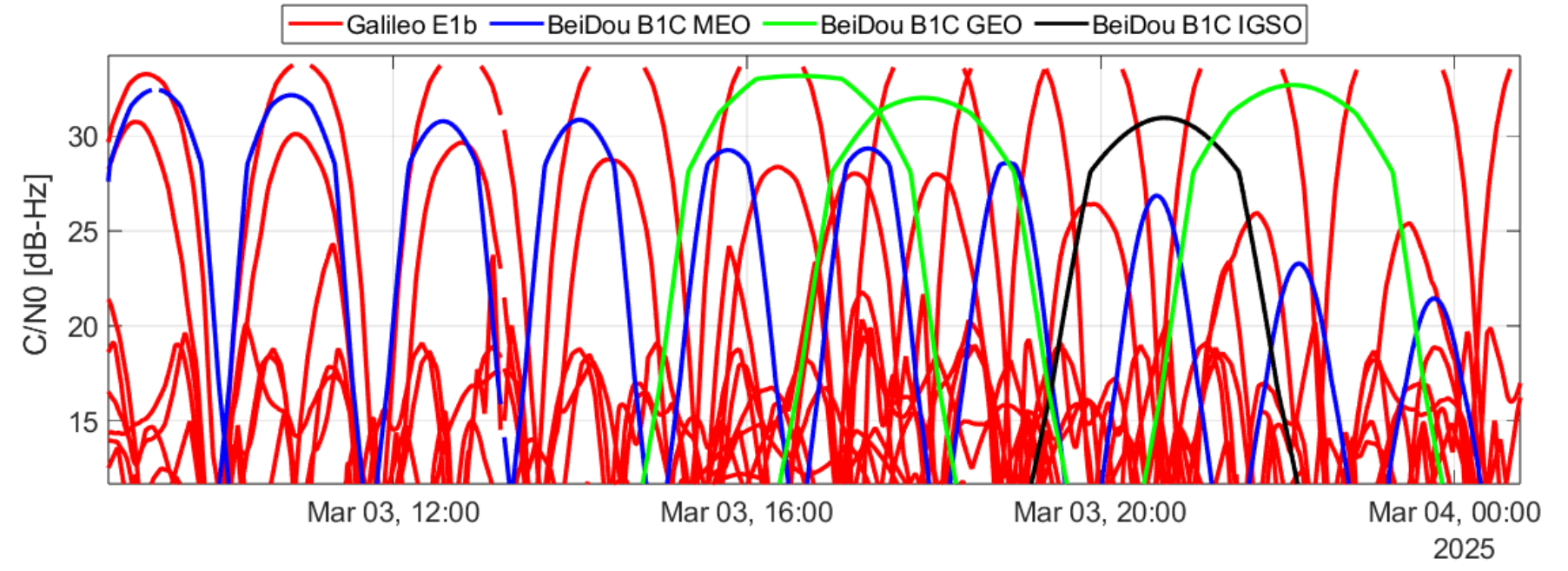}
    \caption{Sample of simulated \gls{C/N0} trends associated to a portion of OP38\_0.}
    \label{fig:pOP38CN0}
\end{figure}

A partial reduction in \gls{GDOP} resulting from the inclusion of additional constellations can be observed in Fig.~\ref{fig:AllGDOP}, which shows the \gls{GDOP} values for all the operations in which the evaluation was possible. As it can be noticed, a wide range of \gls{GDOP} values is observed, reaching a maximum at almost 10,000 and a minimum close to 300. During some operations, including \gls{SBAS} and \gls{RNSS} satellites in the set of satellites used for \gls{GDOP} computation leads to an improvement in \gls{GDOP}. This effect is particularly evident in operations  OP17\_0, OP23\_0, OP37\_0, OP38\_0 and OP76\_0. When GDOP cannot be evaluated without SBAS or RNSS satellites, the contribution to the availability increase becomes evident.
Fig.~\ref{fig:pOP38CN0} illustrates a distinctive operational condition, showing the \gls{C/N0} values generated by the extended \gls{SSV} simulator. During OP38\_0, the Galileo E1B signal exhibits a consistent pattern, as reported in \cite{JournalDiMissione}; signals from \glspl{SV} on orbital plane B progressively fade, while those on plane A remain visible with higher \gls{C/N0} despite increasingly prolonged Earth-shadowing periods. The simulations also show that the \gls{MEO} \glspl{SV} of the BeiDou constellation follow a similar fading pattern, whereas the BeiDou \gls{IGSO} and \gls{GEO} satellites do not. This naturally mitigates the unfortunate phasing phenomena at lunar surface. However, it is important to remark that the use of \gls{SBAS} signals for ranging purposes may require more accurate ephemerides and additional corrections with respect to those broadcast by the \glspl{SV} themselves \cite{kahr2016analysis}.

\begin{figure}[t]
    \centering
    \subfigure[GPS + Galileo\label{fig:AvailGPS_GAL}]{
        \includegraphics[width=0.48\textwidth]{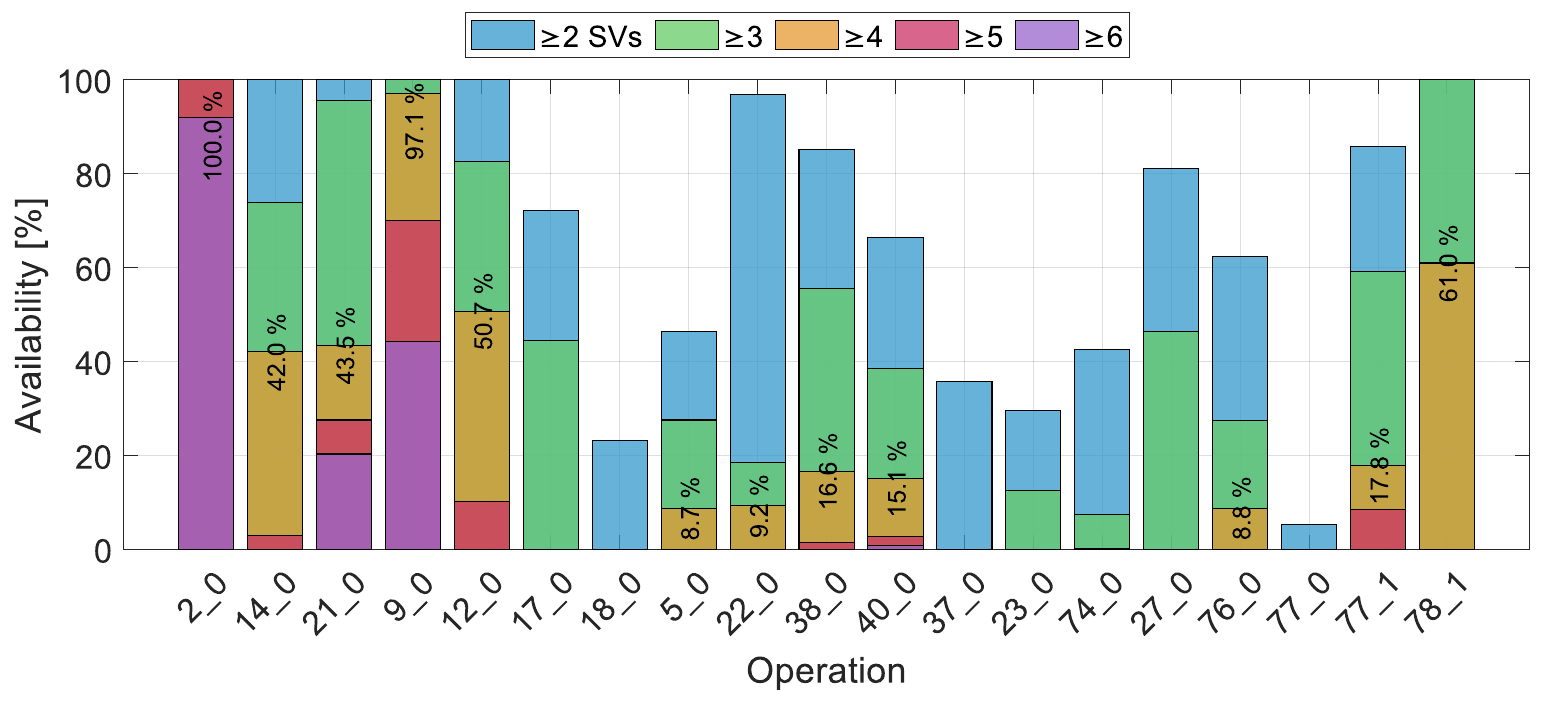}
    }\hfill
    \subfigure[GNSS + RNSS + SBAS\label{fig:AvailRNSS_SBAS}]{
        \includegraphics[width=0.48\textwidth]{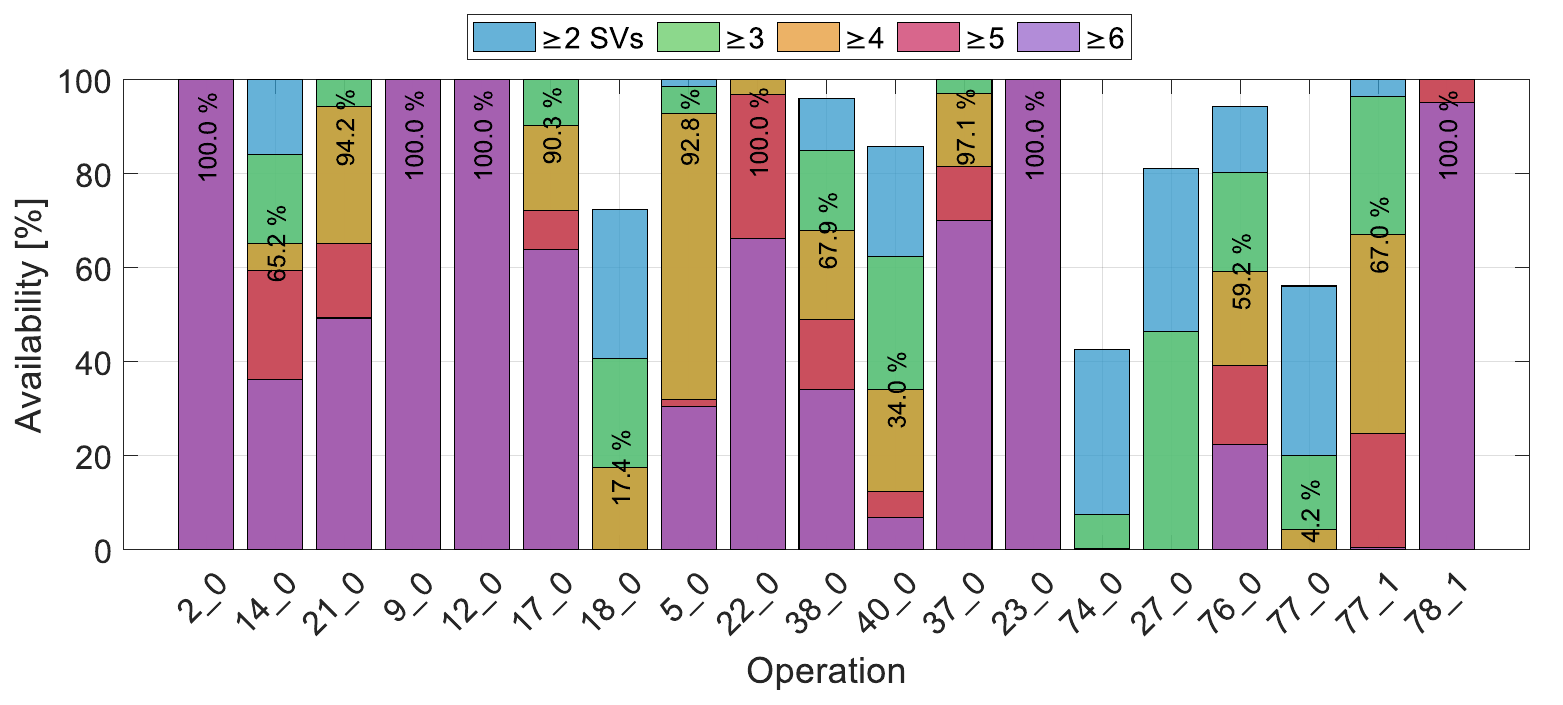}
    }

    \caption{Multi-constellation, multi-frequency availability as the percentage of nominal RTP operation with $n$ signals from $n$ distinct GNSS satellites above the \gls{C/N0} threshold. Percentages are shown for $n \geq 4$.}
    \label{fig:Avail_multi}
\end{figure}

\section{Conclusions}
\label{sec:conclusion}
This work highlights the scientific value of the \gls{LuGRE} \gls{IQS} snapshots for assessing satellite navigation and augmentation systems signal availability in the lunar and cis-lunar environments independently from actual real-time operation of the \gls{LuGRE} payload. Despite significant frequency drift, reduced sampling rates, coarse (4-bit) quantization, and very short snapshot durations, the experimental data demonstrated that signals can be acquired using relatively simple acquisition techniques. 

The experimental results are used to calibrate the extended \gls{SSV} simulator. The simulation of mission operations shows that extending the analysis beyond \gls{GPS} and Galileo to include BeiDou, regional navigation satellite systems such as \gls{IRNSS} (NavIC) and \gls{QZSS}, as well as \gls{SBAS} constellations, leads to a significant improvement in signal availability. In particular, in a dual-frequency multi-constellation configuration, the percentage of total epochs in which at least four satellites are in radiometric visibility increases from 11\% to 46\% with respect to the GPS- and Galileo-only case. Nevertheless, despite this improvement, there remain mission phases in which fewer than four satellites are simultaneously in radiometric visibility. 
In addition to improving availability, processing multiple constellations increases the probability of meeting the minimum visibility required for a-priori-free convergent navigation solutions, allowing on-demand filter initialization without long delays, although high-sensitivity acquisition may still be required. The inclusion of \gls{RNSS} and \gls{SBAS} \gls{GEO} constellations further enhances orbital geometry diversity, mitigating geometry-dependent effects such as orbital-plane–related signal fading observed during specific operational periods (e.g., OP38\_0), and thereby improving availability, at the cost of requiring additional corrections for the \gls{SBAS} \glspl{SV} \cite{kahr2016analysis}.

Overall, the presented results motivate further research on robust high-sensitivity acquisition, frequency-drift compensation, and multi-constellation navigation architectures for future lunar missions. In addition, they highlight the need for more accurate antenna radiation pattern modeling to enhance simulation fidelity, particularly for BeiDou and the \gls{RNSS} and \gls{SBAS} constellations.

\begin{table*}[]
\scriptsize
\centering
\begin{minipage}{0.48\textwidth}
\caption{List of the signals acquired on L1 band. \\ *PRN below the threshold but present on L5,\\
predicted to be close to the radio visibility threshold and with a correlation peak
distinguishable from the noise floor.\\
Excluded from the statistics.}
\label{tab:ACQRESULTS}
\centering
\begin{tabular}{llll}
\hline
\textbf{Operation} & \textbf{Signal} & \textbf{PRN} & \textbf{C/N\textsubscript{0} [dB-Hz]} \\
\hline
\multicolumn{4}{c}{\textbf{Commissioning and Transit}} \\
\hline
\multirow{4}{*}{OP2\_0}
 & GPS C/A & 18  & 39.7 \\
 & GPS C/A & 24  & 37.4 \\
 & E1B     & 36  & 33.8 \\
 & B1C     & 34  & 33.6 \\
\midrule
\multirow{3}{*}{OP5\_0}
 & GPS C/A & 28  & 34.6 \\
 & SBAS L1 - BDSBAS  & 130 & 30.6  \\
 & SBAS L1 - SouthPAN & 122 & 34 \\
\midrule
\multirow{1}{*}{OP12\_0}
 & QZSS L1 & 3   & 36 \\
\midrule
\multirow{3}{*}{OP14\_0}
 & B1C     & 36  & 35 \\
 & SBAS L1 - WAAS    & 133 & 37 \\
 & SBAS L1 - WAAS    & 135 & 36.8 \\
\midrule
\multirow{4}{*}{OP17\_0}
 & GPS C/A & 11  & 34.3 \\
 & GPS C/A & 30  & 31.8 \\
 & SBAS L1 - BDSBAS  & 144 & 35.8 \\
 & SBAS L1 - KASS    & 134 & 30.1 \\
\midrule
\multirow{1}{*}{OP18\_0} 
 & SBAS L1 - WAAS  & 131 & 34.2 \\
 \midrule
 \multirow{3}{*}{OP21\_0}
 & GPS C/A & 24  & 35.8 \\
 & SBAS L1 - WAAS    & 131 & 39.8 \\
 & E1B     & 11  & 32.4 \\
 \midrule
  \multirow{3}{*}{OP22\_0}
 & GPS C/A & 31  & 34\\
 & QZSS L1 & 3  & 37.7 \\
 & SBAS L1 - MSAS    & 137 & 30.9 \\
\midrule
\multicolumn{4}{c}{\textbf{Lunar Orbit}} \\
\hline
\multirow{4}{*}{OP23\_0}
 & B1C     & 39  & 33 \\
 & E1B     & 33  & 32.2\\
 & NavIC L1 & 10 & 32.2\\
 & SBAS L1 - BDSBAS  & 130 & 32.4 \\
\midrule
\multirow{2}{*}{OP32\_0}
 & B1C     & 34  & 34.2 \\
 & B1C & 50* & -- \\
\midrule
\multirow{5}{*}{OP37\_0}
 & B1C     & 39  & 29.1 \\
 & QZSS L1  & 3* & -- \\
 & QZSS L1  & 7* & -- \\
 & SBAS L1 - BDSBAS  & 143 & 37.3 \\
 & SBAS L1 - MSAS    & 137 & 32 \\
\midrule
\multicolumn{4}{c}{\textbf{Lunar Surface}} \\
\hline
\multirow{2}{*}{OP40\_0}
 & GPS C/A & 25  & 31.1 \\
 & E1B     & 31  & 31.3 \\
\midrule
\multirow{1}{*}{OP73\_0}
 & E1B     & 33  & 32.7 \\
\midrule
\multirow{3}{*}{OP74\_0}
 & B1C     & 26  & 30.8 \\
 & GPS C/A & 31  & 33.2 \\
 & SBAS L1 - EGNOS  & 136 & 35 \\
\midrule
\multirow{2}{*}{OP76\_0}
 & GPS C/A & 31  & 32.2 \\
 & E1B     & 26  & 30.3 \\
\midrule
\multirow{3}{*}{OP77\_0}
 & B1C   & 45  & 29.3 \\
 & E1B   & 13  & 32.6 \\
 & SBAS L1 - AL-SBAS & 148 & 31.3 \\
\midrule
\multirow{1}{*}{OP77\_1}
 & E1B & 33 & 28.9 \\

\midrule
\multirow{5}{*}{OP78\_0}
 & B1C & 37  & 29.2 \\
 & GPS C/A & 12 & 34.1 \\
 & SBAS L1 - WAAS & 131 & 38.8 \\
 & SBAS L1 - WAAS & 133 & 33.8 \\
 & SBAS L1 - WAAS & 135 & 36.5 \\
\midrule
\multirow{5}{*}{OP78\_1}
 & B1C & 37 & 32 \\
 & GPS C/A & 12 & 35 \\
 & SBAS L1 - WAAS & 131 & 35.3 \\
 & SBAS L1 - WAAS & 133 & 36 \\
 & SBAS L1 - WAAS & 135 & 35.3 \\
\midrule
\hline

\end{tabular}
\vspace{3pt}

\end{minipage}
\hfill
\begin{minipage}{0.48\textwidth}
\caption{List of the signals acquired on L5 band. }
\label{tab:ACQRESULTSL5}
\centering
\begin{tabular}{llll}
\hline
\textbf{Operation} & \textbf{Signal} & \textbf{PRN} & \textbf{C/N\textsubscript{0} [dB-Hz]}\\ 
\hline
\multicolumn{4}{c}{\textbf{Commissioning and Transit}} \\
\hline
\multirow{11}{*}{OP9\_0}
 & E5aI            & 26           & 38.8 \\
 & B2aD            & 34           & 34.1 \\
 & QZSS L5I        & 3            & 30.8 \\
 & QZSS L5I        & 4            & 30.5 \\
 & QZSS L5I        & 7            & 35.1 \\
 & NavIC L5        & 4            & 28.5 \\
 & NavIC L5        & 5            & 32.2 \\
 & NavIC L5        & 7            & 41.7 \\
 & SBAS L5Q - BDSBAS & 130        & 43.2 \\
 & SBAS L5Q - QZSS  & 185        & 39.6 \\
 & SBAS L5Q - QZSS  & 186        & 39.8 \\
\midrule
\multicolumn{4}{c}{\textbf{Lunar Orbit}} \\
\hline
\multirow{3}{*}{OP27\_0}
 & L5Q             & 23           & 34.8 \\
 & L5Q             & 26           & 33.7 \\
 & NavIC L5        & 6            & 30.7 \\
\midrule
\multirow{3}{*}{OP31\_0}
 & L5Q             & 23           & 35.9 \\
 & B2aD            & 29           & 33.9 \\
 & B2aD            & 48           & 35.5 \\
\midrule
\multirow{5}{*}{OP32\_0}
 & B2aD            & 34           & 37 \\
 & B2aD            & 50           & 32.9 \\
 & B2aD            & 40           & 30.7 \\
 & B2aD            & 43           & 29.8 \\
 & NavIC L5        & 5            & 28.8 \\
\midrule
\multirow{8}{*}{OP37\_0}
 & B2aD            & 39           & 36.4 \\
 & B2aD            & 44           & 30.4 \\
 & QZSS L5I        & 7            & 30.8 \\
 & QZSS L5I        & 3            & 28 \\
 & NavIC L5        & 5            & 28.8 \\
 & NavIC L5        & 10           & 29.2 \\
 & SBAS L5Q - BDSBAS & 143        & 40.2 \\
 & SBAS L5Q - QZSS  & 185        & 36.4 \\
\midrule
\multicolumn{4}{c}{\textbf{Lunar Surface}} \\
\hline
\multirow{1}{*}{OP40\_0}
 & E5aI            & 31           & 33.4 \\
\midrule
\multirow{2}{*}{OP73\_0}
 & L5Q             & 24           & 30.7 \\
 & E5aI            & 33           & 34.8 \\
\midrule
\multirow{1}{*}{OP74\_0}
 & B2aD            & 26           & 34.9 \\
\midrule
\multirow{1}{*}{OP76\_0}
 & E5aI            & 26           & 31.3 \\
\midrule
\hline
\end{tabular}

\end{minipage}
\end{table*}

\section*{Acknowledgment}
This study was funded within the contract n. 2021-26-HH.0 between Agenzia Spaziale Italiana and Politecnico di Torino ”Attività di Ricerca e Sviluppo inerente alla Navigazione GNSS nello Space volume Terra/Luna nell’ambito del Lunar GNSS Receiver Experiment”. 

This publication is also  part of the project PNRR-NGEU which has received funding from the MUR – DM 630/2024.

\bibliography{references}

\begin{IEEEbiography}[{\includegraphics[width=1in,height=1.25in,clip,keepaspectratio]{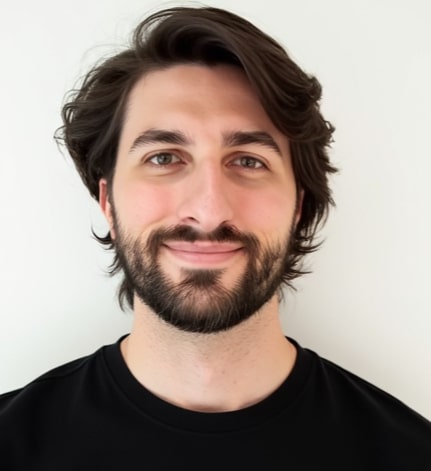}}]{Lorenzo Sciacca}{\space}(Graduate Student Member, IEEE) received his M.Sc. in Communications and Computer Network Engineering from Politecnico di Torino, Italy, in 2024. He is currently pursuing a Ph.D. at the Department of Electronics and Telecommunications, Politecnico di Torino, as part of the Navigation Signal Analysis and Simulation (NavSAS) group. His research focuses on developing high-sensitivity, environment-aware GNSS receivers to improve navigation in challenging conditions, including lunar and cis-lunar environments. He has been involved in the Lunar
GNSS Receiver Experiment (LuGRE) since 2025.
\end{IEEEbiography}%

\begin{IEEEbiography}[{\includegraphics[width=1in,height=1.25in,clip,keepaspectratio]{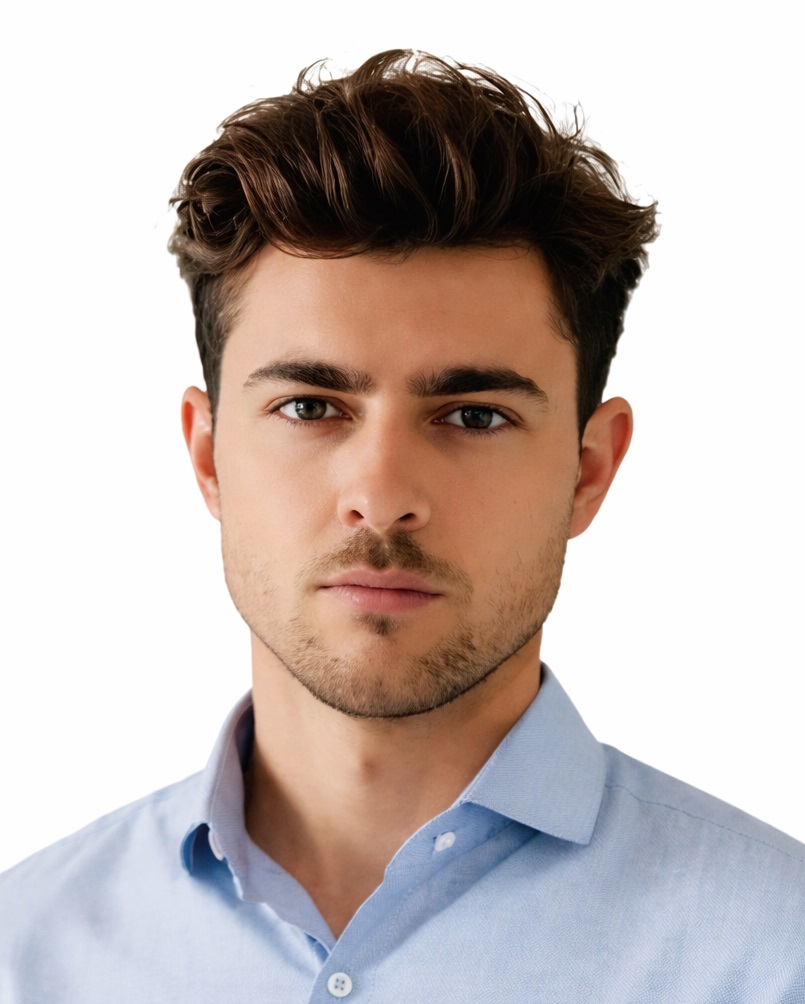}}]{Alex Minetto}{\space} (Member, IEEE) received the B.Sc. (2013), M.Sc. (2015), and Ph.D. with honors (2020) in telecommunications and electrical engineering from Politecnico di Torino, where he is now Assistant Professor in the Department of Electronics and Telecommunications and a member of the NavSAS research group. His work focuses on advanced GNSS signal analysis, receiver design, and Bayesian state estimation. He has contributed to ESA and EUSPA projects and since 2021 has been involved in the Lunar GNSS Receiver Experiment (LuGRE) as a mission's Science Team member.
\end{IEEEbiography}

\begin{IEEEbiography}[{\includegraphics[width=1in,height=1.25in,clip,keepaspectratio]{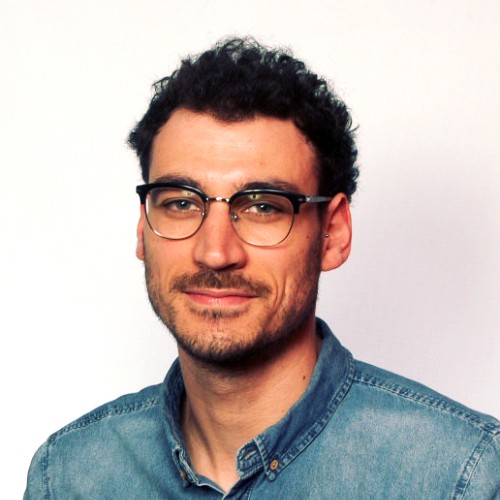}}]{Andrea Nardin} (Member, IEEE) received the M.Sc. (2018) and Ph.D. with honors (2023) in telecommunications and electrical engineering from Politecnico di Torino, where he is now Assistant Professor in the Department of Electronics and Telecommunications and a member of the NavSAS research group. In 2021, he was a Visiting Doctoral Researcher at Northeastern University, Boston. His research interests include signal processing architectures and signal design for GNSS and next-generation PNT systems.
\end{IEEEbiography}

\begin{IEEEbiography}[{\includegraphics[width=1in,height=1.25in,clip,keepaspectratio]{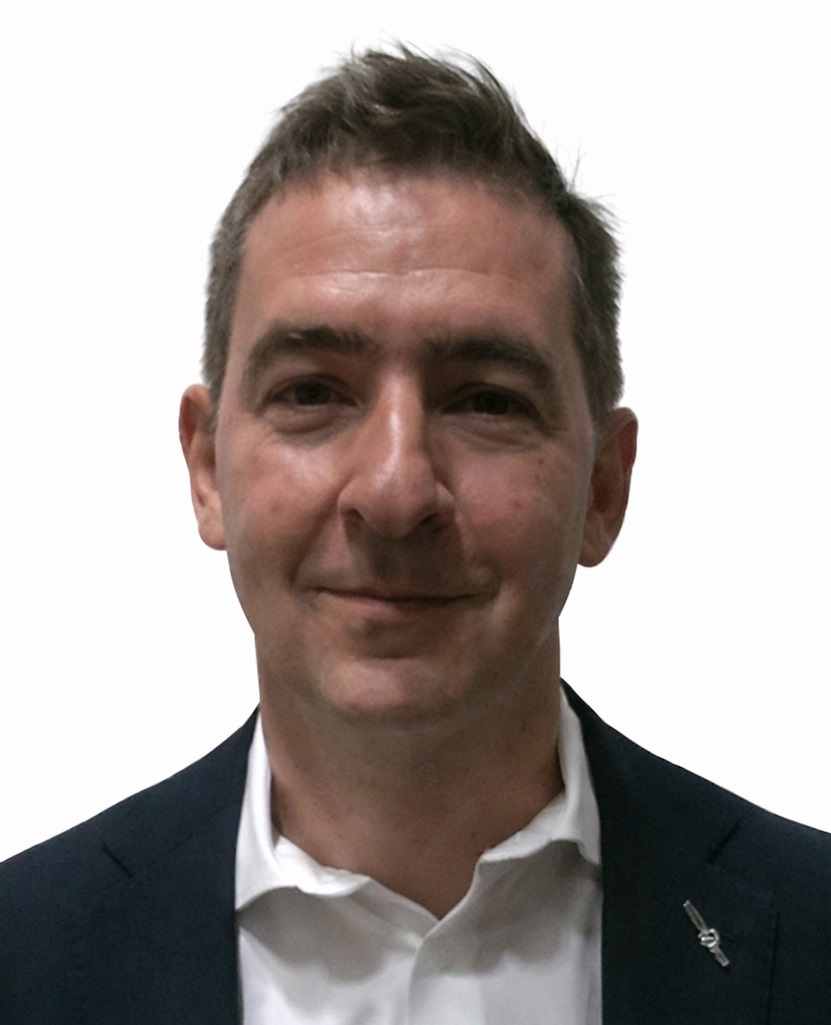}}]{Fabio Dovis} (Member, IEEE) received the M.Sc. (1996) and Ph.D. (2000) from Politecnico di Torino, where he is now Full Professor in the Department of Electronics and Telecommunications and coordinator of the NavSAS Research Group. His research focuses on GNSS receiver design, advanced signal processing for interference and multipath mitigation, and ionospheric monitoring. He has extensive experience in international projects, collaborations with industry and research institutions, and serves on the IEEE AESS Navigation Systems Panel.
\end{IEEEbiography}

\begin{IEEEbiography}[{\includegraphics[width=1in,height=1.25in,clip,keepaspectratio]{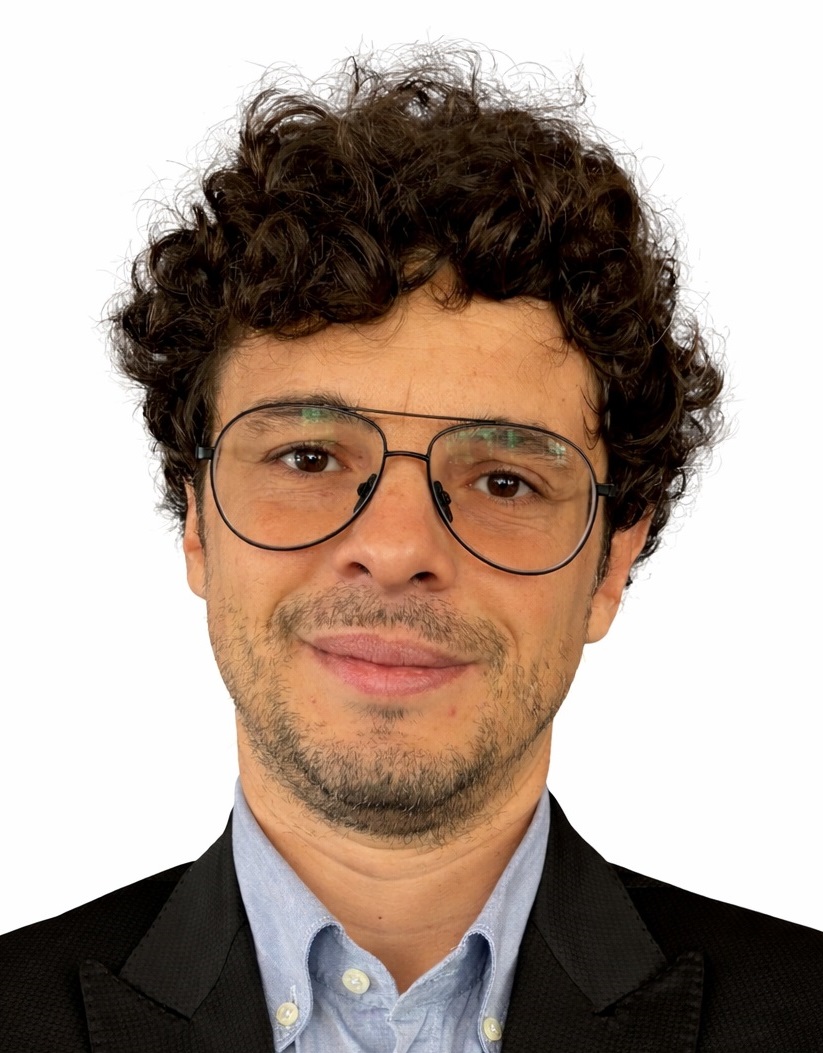}}]{Luca Canzian} joined Qascom in 2015 and he is currently leading the R\&D domain area. He has worked in several projects with ESA, ASI, NASA, the European Commission and Industry. His main expertise involves ground-based and space-based location systems, detection and location of interference signals, navigation using signals of opportunity, lunar navigation, inertial navigation systems, orbit determination techniques, GNSS authentication and anti-spoofing techniques. He holds a MSc degree and a PhD in Electrical Engineering from University of Padova (Italy).
\end{IEEEbiography}

\begin{IEEEbiography}[{\includegraphics[width=1in,height=1.25in,clip,keepaspectratio]{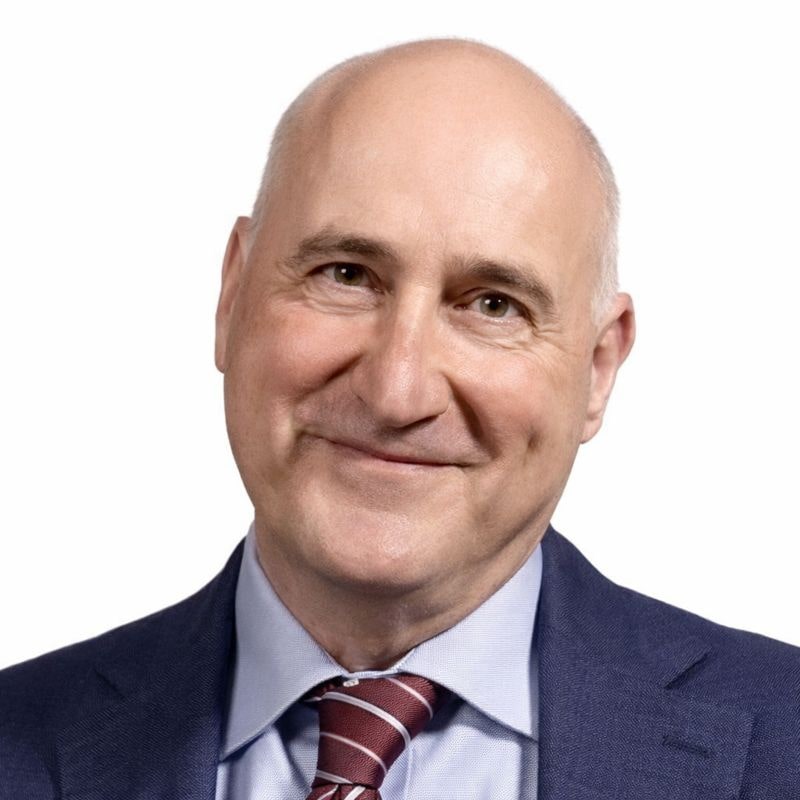}}]{Mario Musmeci} received a degree in physics from the University of Rome “La Sapienza.” He has worked with major Italian space companies, ESA, and the European Commission on the Galileo program, and co-founded a start-up exploiting a patent based on Galileo signals for reliable traceability of “Made in Italy” products. Since 2016, he has been with ASI’s Programs Directorate, serving as technical manager for LuGRE on NASA’s Firefly Blue Ghost mission, demonstrating GNSS reception on the lunar surface.
\end{IEEEbiography}

\begin{IEEEbiography}[{\includegraphics[width=1in,height=1.25in,clip,keepaspectratio]{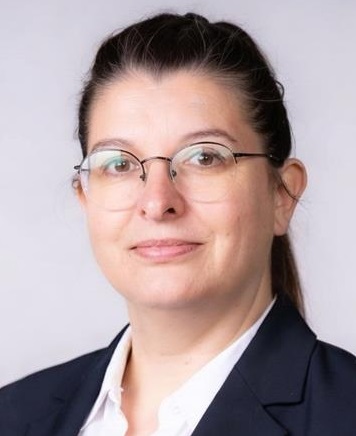}}]{Claudia Facchinetti} is a researcher and technologist at the Italian Space Agency (ASI), specializing in Earth observation and space navigation technologies. She holds a Ph.D. in Space Science and Technology and has contributed to numerous scientific projects and space missions, including studies on SAR radar and hyperspectral sensors. She is a key member of the LuGRE (Lunar GNSS Receiver Experiment) team, a joint NASA-ASI initiative that successfully demonstrated GNSS-based navigation on the Moon during the Firefly Blue Ghost Mission 1 in 2025. Additionally, she is responsible for developing Earth observation missions at ASI.
\end{IEEEbiography}

\begin{IEEEbiography}[{\includegraphics[width=1in,height=1.25in,clip,keepaspectratio]{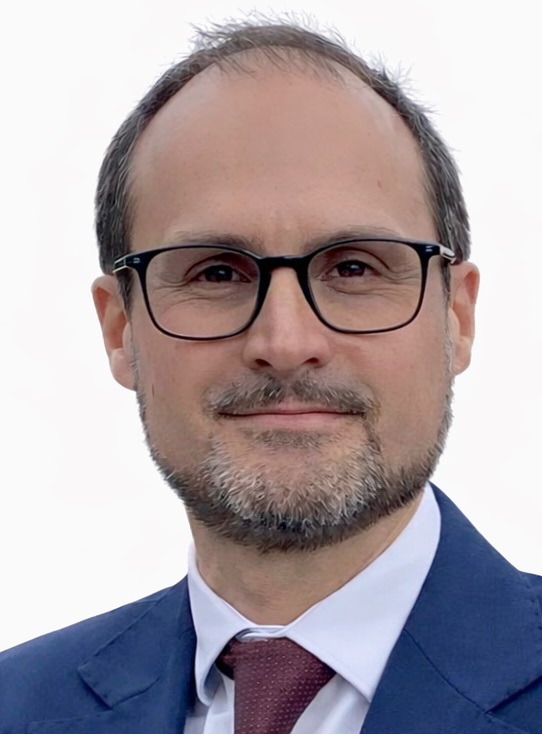}}]{Giancarlo Varacalli} received the degree in Electronic Engineering and holds a Master’s degree in Space Systems Engineering from Delft University of Technology. He is currently Head of the Telecommunications and Navigation Department at the Italian Space Agency (ASI). He has held several technical and managerial positions within ASI and serves as Italy’s delegate to ESA programme boards in satellite communications and navigation. His professional interests include satellite telecommunications, navigation systems, and space system engineering.
\end{IEEEbiography}

\bibliographystyle{ieeetr}  

\end{document}